\documentclass[epj]{svjour}
\usepackage{graphicx}
\usepackage{amsmath,amssymb}

\let\I\i
\def\i{\mathrm{i}}

\def\d{\mathrm{d}}
\def\half{{\textstyle{1\over2}}}
\def\thalf{{\textstyle{3\over2}}}
\def\h{{\scriptscriptstyle{1\over2}}}
\def\th{{\scriptscriptstyle{3\over2}}}

\def\vec#1{\mbox{\boldmath$#1$}}
\def\CG#1#2#3#4#5#6{C^{#5#6}_{#1#2#3#4}}
\def\DS{{\Delta^*}}
\def\RS{{R^*}}

\begin{document}

\title{Roper resonances in chiral quark models}

\author{%
B.~Golli\inst{1}
\and
S.~\v{S}irca\inst{2}}

\institute{%
Faculty of Education,
              University of Ljubljana and J.~Stefan Institute,
              1000 Ljubljana, Slovenia
\and
Faculty of Mathematics and Physics,
              University of Ljubljana and J.~Stefan Institute,
              1000 Ljubljana, Slovenia}

\date{\today}

\abstract{%
We derive a method to calculate the multi-channel K matrix
applicable to a broad class of models in which 
mesons linearly couple to the quark core.
The method is used to calculate pion scattering
amplitudes in the energy region of low-lying P11 and P33 
resonances.
A good agreement with experimental data is achieved if 
in addition to the elastic channel we include the $\pi\Delta$ 
and $\sigma N$ ($\sigma\Delta$) channels where the $\sigma$-meson 
models the correlated two-pion decay.
We solve the integral equation for the K matrix in the
approximation of separable kernels; it yields a sizable 
increase of the widths of the $\Delta(1232)$ and the $N(1440)$
resonances compared to the bare quark values.}


\maketitle

\section{\label{sec:intro}Introduction}

Among the low-lying nucleon excitations, the P11 Roper resonance 
$N(1440)$ as well as its counterpart in the P33 partial wave, 
the $\Delta(1600)$, play a special role due to their 
relatively low masses and due to the rather peculiar 
behavior of the scattering and electro-excitation amplitudes.
The constituent quark model (CQM) in which the excited states 
are treated as bound states yields partial decay widths which 
are generally considerably smaller than the experimental values
\cite{melde05,faessler02,koniuk8280,forsyth83,isgur7978,capstick,%
theussl,stancu88} unless all widths are scaled
by a factor of 2--3 in order to fit the experimental width
of the $\Delta(1232)$.  This indicates that the structure of 
the Roper can not be explained by a simple excitation of the 
quark core (like most of the other low-lying states) and that 
other degrees of freedom need to be included.

Several attempts to understand and explain the nature of the 
Roper resonance have been proposed and discussed.
They include the study of the Roper resonance on the lattice 
\cite{burch05,guadagnoli0504,mathur05,sasaki0205}, 
investigations in different quark models,
like the models based on $qqqq\overline{q}$ configurations
\cite{dillig07,faessler07,riska06} or those incorporating the 
meson cloud consisting primarily of pions and $\sigma$-mesons
\cite{Pedro,afnan89,cooper86,Veit86,kalbermann84,Suzuki83,Rinat82,%
thomas8180},
as well as in models with hybrid ($qqqg$ or glueball)
configurations \cite{Li92,broRPA}.  
Pion-nucleon scattering in the region of the Roper resonance has 
also been studied in the framework of chiral perturbation theory
\cite{oset02,fettes01,jensen97}.
The scattering amplitudes and the decay rates
have been well established in certain phenomenological approaches, 
for example in energy-independent partial-wave analysis of
$\pi N\rightarrow \pi\pi N$ scattering
\cite{manley92,cutkosky90,longacre77}, energy-dependent analysis 
\cite{SAID06,vrana00}, model-independent analysis 
\cite{ceci06}, 
as well as in dynamical coupled-channel models
\cite{satolee07,julia-diaz07}
and in effective-Lagrangian models (see, for example,
\cite{kamano060504,penner02,speth00,morsch99,gridnev99,feuster99,%
speth98,gross93}).
Such models typically contain a large number of ingredients 
({\it e.g.\/} coupling constants, cut-offs, meson and baryon 
states) in order to parameterize the observed scattering  
(or electro-production) amplitudes and to simultaneously fit 
many measured observables.
However, because of the multitude of model ingredients,
it is usually difficult to determine which  degrees of 
freedom are truly relevant for a particular resonance
and to obtain insight into its quark structure.
Furthermore, a similar level of agreement with data can be 
achieved using rather disparate sets of parameters;
using results of quark-model calculations may offer an 
important guidance in particular to the choice of the form
factors of resonant states.

The first aim of the present work is to develop a method which 
would allow solutions obtained in a broad class of quark-model 
calculations using bound-state boundary conditions to be 
incorporated into a dynamical framework that could be used to 
predict observables measured in meson scattering experiments 
as well as in electro-weak production of mesons on the nucleon.
The interplay of different baryonic and mesonic degrees of freedom
in such processes can in many cases considerably alter the
results obtained by simply calculating the transition
form-factors of the baryons in the underlying quark model.

In addition, our goal is to set up a computational scheme 
beyond the simple approximation for the K matrix and connect
the observables extracted in our analysis of the relevant
degrees of freedom to the predictions of the underlying
quark model, which would eventually give us a chance
to discriminate between different models.

Our method is not intended to be a competitive approach 
to those using effective Lagrangians.  Rather, the method 
complements them by establishing a link between the 
effective models and the underlying baryon structure. 
Its main distinctions with respect to the (non-dynamical) 
effective Lagrangian approaches are:
(i) baryons are treated as composite particles from the 
very beginning; the strong and electro-weak form-factors 
are derived from baryon internal structure and not 
inserted a posteriori; as a consequence the method 
introduces a much smaller number of free parameters;
(ii) the meson cloud around baryons is included in a 
consistent way also in the asymptotic states;
(iii) through the Kohn variational principle for 
the K matrix it allows one to determine the internal 
structure of the baryon resonances which may substantially 
differ from the structure determined in the calculations 
using bound-state boundary conditions.
On the other hand, because of the complex structure of 
the baryons in the model, we are not able to treat a very 
large set of ingredients. Consequently, the approach in 
its present form cannot be used at higher energies where 
many channels open.

In several aspects the present method is similar to the 
method of solving the integral equation for the T matrix 
described in \cite{satolee07} and used in \cite{julia-diaz07}.
In both approaches the coupled-channel equations are 
solved beyond the usual Born approximation including the 
effect of the meson cloud in the baryon; as a consequence, 
in both cases the expression of the T matrix (K matrix)
involves the physical nucleon rather than a structureless
particle of effective Lagrangian approaches.
The starting points are however different: the approach of
\cite{satolee07} starts from baryon-meson and meson-meson
Lagrangians and using the unitary transformation method 
derives a model Hamiltonian with dynamically generated 
baryon-meson vertices which can be related to the vertices 
calculated in quark models.
Our approach is complementary in that it starts from the
quark-model type calculation and extends the many-body
wave-function such that it properly includes the
asymptotic boundary conditions of a particular channel.
The defficiency of our approach in the present form
is the lack of meson self-interactions.

Our future plan is to extend the approach to include 
meson photo- and electro-production which requires 
the inclusion of new channels.
In order not to deal with an excessively large number of
ingredients in that case, the aim of the present analysis 
is also to reveal which degrees of freedom are most relevant 
for an explanation of the structure and the dynamics of the 
low-lying resonances in the P11 and P33 partial waves and 
to pin down the parameters governing these resonances.

In our previous work~\cite{EPJA} we have developed a method to 
calculate the scattering and the pion electro-production amplitudes 
in models in which the pions linearly couple to the quark core.
In such models it is possible to find 
the exact expression for the T matrix as well as for the K matrix 
without explicitly specifying the form of the asymptotic states.
The method has been successfully applied to the calculation 
of the phase shift and the pion electro-production amplitudes for 
the P33 partial wave by considering only the elastic channel.
In this paper we extend the method to calculate directly 
the multi-channel K-matrix.
The resulting matrix is symmetric and real, thus ensuring unitarity.
Since we are now considering processes at relatively higher energies
we generalize the approach by adopting fully relativistic kinematics.

The first attempt to treat dynamically the resonant pion scattering
in a quark model with excitations of the quark core has been
done in the framework of the Cloudy Bag Model (CBM). 
For the P33 partial wave \cite{Veit86,thomas8180} the experimental 
phase shift in the energy range of the $\Delta(1232)$ resonance has 
been well reproduced.
In the $N(1440)$ case the width of the resonance has been considerably 
underestimated  \cite{afnan89,Veit86,Rinat82} except in the case of 
the coupled-channel calculation \cite{Suzuki83} using the K-matrix 
approach involving the $\pi\Delta$ and $\pi$-Roper inelastic 
channels in the static approximation. 
The latter approach is similar to the me\-thod of our 
sect.~\ref{sec:Born} except that we use relativistic kinematics.

In the present work we restrict ourselves to a class of
Hamiltonians with linear meson-baryon coupling,
neglecting meson self-interaction.
We are particularly interested in the interplay
of the degrees of freedom involving the $\Delta$ isobar and the
$\sigma$-meson which we expect to play the dominant role in the
dynamics of the Roper resonances; in this respect our work 
is similar to that of Ref.~\cite{speth00,krewald06}.
We were, however, not able to confirm the most intriguing
conclusion found in \cite{speth00} that the formation of the
$N(1440)$ resonance can be explained without introducing
a bare Roper state.

In sect.~\ref{sec:model} we introduce a general form of models 
in which mesons linearly couple to the quark core.
The input to our computational scheme are the matrix elements
of the meson interaction between the quark states.
We construct the K matrix and the corresponding principal-value 
states which incorporate in a systematic way the many-quark 
quasi-bound states describing the excitations of the quark core.
We assume that the decay into two pions which dominates the
inelastic processes in the lower energy region proceeds through
a two-body decay involving either an unstable isobar
which in turn decays into the nucleon and the second pion
or an unstable meson and the nucleon.

In sect.~\ref{sec:integral} we show how a quasi-bound state 
calculated in the underlying quark model can be inserted
in a state that obeys proper scattering boundary conditions.
We derive a set of coupled integral equations for the
meson amplitudes and the parameters governing the strength
of the quasi-bound states which are responsible for the
resonant parts of the scattering amplitudes.
We first solve the set in a simple approximation equivalent 
to the so-called Born approximation for the K matrix widely 
used in various multi-channel approaches.
In order to find the solution beyond the Born approximation
we introduce an approximation which makes the kernels separable, 
however, ensuring that they reduce to the exact form when 
evaluated on-shell, as well as preserving the proper symmetries 
of the K matrix.

The results of our analysis using the quark wave-func\-tions from 
the Cloudy Bag Model are presented in sect.~\ref{sec:results}.
We analyze to what extent various approximations are able to 
explain the resonant behavior of the P11 and P33 partial amplitudes.
We stress the important role of the background terms originating from 
the crossed ($u$-channel) processes, and show that the inclusion of the 
$\sigma N$ channel considerably improves the results in the P11 case.
The meson-baryon coupling constants needed to fit the scattering 
amplitudes in the Born approximation for the K matrix are generally 
considerably larger than those predicted by the quark model.
Solving the integral equation for the K matrix beyond the Born 
approximation we show that the resulting dressing of the vertices 
as well as the influence of the neighboring resonances 
can explain the strong enhancement of the coupling constants.


\section{\label{sec:model}Coupled-channel K-matrix formalism}

\subsection{\label{sub:Kmatrix}The K matrix for models with linear 
meson-quark coupling}

We consider here a rather broad class of models in 
which the mesons linearly couple to the three-quark core.
The part of the Hamiltonian referring to mesons can be written as
\begin{eqnarray}
 H_\pi &=& 
  \int\d k \sum_{lmt}\left\{\omega_k\,a^\dagger_{lmt}(k)a_{lmt}(k)
\right. \nonumber\\ && \left.
    + \left[V_{lmt}(k) a_{lmt}(k) 
        + V_{lmt}^\dagger(k)\,a^\dagger_{lmt}(k)\right] \right\},
\label{Hpi}
\end{eqnarray}
where
$a^\dagger_{lmt}(k)$ is the creation operator for a $l$-wave
meson with the third component of spin $m$ and -- in the case 
of isovector mesons -- the third component of isospin $t$.
In the case of the $p$-wave pions, the source can be cast 
in the form
\begin{equation}
      V_{mt}(k) = -v(k)\sum_{i=1}^3 \sigma_m^i\tau_t^i\,,
\label{Vmt}
\end{equation}
with $v(k)$ depending on a particular quark model and containing 
the information about the underlying quark structure.
(We omit the index $l$ from the pion operators.)
We assume that the interaction $V(k)$ can generate bare quark 
states with quantum numbers different from the ground state 
by flipping the spin and isospin of the quarks, and
furthermore, excite quarks to higher spatial states.
In particular,  the state with the flipped spins and isospin
corresponds to the $\Delta(1232)$ isobar, while exciting one 
quark from the 1s to the 2s state generates an excited three-quark 
state associated with the Roper resonances $N(1440)$ 
or $\Delta(1600)$.

Chew and Low \cite{ChewLow} considered a model similar to (\ref{Hpi}) 
except that they did not allow for excitations of the nucleon core.
They showed that the T~matrix for $\pi N$ scattering was proportional 
to $\langle\Psi^{(-)}(W)|V_{mt}(k)|\Psi_N\rangle$,
where $\Psi^{(-)}(W)$ were the incoming states.
In general, the corresponding formula for the K~matrix cannot be
written in such a simple form.
However, in the $JT$ basis, in which the K and T~matrices are 
diagonal, it is possible to express the K~matrix as
\begin{equation}
   K_{\pi N\pi N}^{JT}(k,k_0) =  -\pi\sqrt{\omega_k E_N\over k\, W}
      \langle\Psi^N_{JT}(W)||V(k)||\Psi_N\rangle\,.
\label{KCL}
\end{equation}
The corresponding principal-value (PV) state obeys a similar 
equation as the in- and out-going states in the Chew-Low model
and takes the form:
\begin{eqnarray}
  |\Psi^N_{JT}(W)\rangle 
&=& 
 \mathcal{N}_0
  \biggl\{
   \left[a^\dagger(k_0)|\Psi_N\rangle\right]^{JT}
\biggr. \nonumber\\ && \biggl.
  - {\mathcal{P}\over H-W}\,
  \left[V(k_0)|\Psi_N\rangle\right]^{JT}
    \biggr\}\,,
\label{PVN}
\end{eqnarray}
where $[\kern3pt]^{JT}$ denotes coupling to good $J$ and $T$, and
\begin{equation}
\mathcal{N}_0 = \sqrt{\omega_0E_N\over k_0 W}\,.
\label{calN0}
\end{equation}
We work in the center-of-mass system, $W$ is the invariant energy 
of the system, $\omega_0$ and $k_0= \sqrt{\omega_0^2 - m_\pi^2}$ 
are the energy and momentum of the pion, 
and $E_N$ is the nucleon energy:
$$
   \omega_0 = W - E_N = {W^2 - M_N^2 + m_\pi^2\over 2W}\,,
\quad
E_N = \sqrt{M_N^2 + k_0^2}\,.
$$

It is worthwhile to notice that a PV state is a superposition of 
incoming and outgoing waves ({\it i.e.\/} can be regarded as a 
standing wave); the in- and out-states become meaningful only 
when the T and the S matrix are constructed from the K matrix:
\begin{equation}
    T = {K\over (1-\i K)}\,,
\qquad
    S = 1 + 2\i T\,.
\label{TSmatrix}
\end{equation}

\subsection{Channels including the pion and unstable isobars}

We now extend the above formulas to the multichannel case.
We apply the usual approach used in phenomenological analyses 
({\it e.g.\/} \cite{manley92}) in which it is assumed that 
the two-pion decay proceeds through some intermediate unstable 
particle, either a meson or a baryon.

Let us first consider the situation in which the intermediate 
particle is a baryon, $B$, ({\it e.g.\/} the $\Delta$ isobar which 
dominates at low energies) which in turn decays into the 
nucleon and the second pion.
The decays into the nucleon and an unstable meson will 
be treated in the next subsection.

In analogy with (\ref{PVN}) we introduce the PV state corresponding 
to the $\pi B$ channel as
\begin{eqnarray}
 |\Psi^B_{JT}(W,M)\rangle &=& \mathcal{N}_1
  \biggl\{
   \left[a^\dagger(k_1)|\widetilde{\Psi}_B(M)\rangle\right]^{JT}
\biggr. \nonumber\\ && \biggl.
  - {\mathcal{P}\over H-W}\,
  \left[V(k_1)|\widetilde{\Psi}_B(M)\rangle\right]^{JT}
    \biggr\}\,.
\label{PVB}
\end{eqnarray}
Here $k_1$ is the momentum of the first pion; the energy $E$ 
of the isobar $B$ and the energy of the first pion are related 
through
\begin{equation}
E  = {W^2 + M^2 - m_\pi^2 \over 2W}\,,
\qquad 
\omega_1 = W - E\,,
\label{kinema1}
\end{equation}
where $M$ is the mass of $B$, {\it i.e.\/} 
the invariant mass of the nucleon and the second pion, and
\begin{equation}
  \mathcal{N}_1 = \sqrt{\omega_1E\over k_1 W}\,.
\label{calN1}
\end{equation}

The normalization of the PV states (\ref{PVN}) and (\ref{PVB}) 
states is dictated by their respective first terms which 
represent a free pion and the nucleon or the $B$ isobar.
The states $\Psi^\alpha$, $\alpha=N, (B,M), \ldots$ 
are then normalized as
$$
   \langle\Psi^\alpha(W) |
                     \Psi^\beta(W')\rangle
  = \delta(W-W') \left[\delta_{\alpha,\beta} 
  + {\mathbf{K}^2}_{\alpha,\beta}\right]
$$
(see {\it e.g.\/} \cite{Newton}, eq.~(7.28)).
In the case of the $\pi N$ channel $\alpha=N$ is a single index, 
while in the $\pi B$ channel it includes also the invariant 
mass $M$, so $\delta_{\alpha,\alpha}$ should be interpreted 
as $\delta(M-M')$.
The PV states are not orthonormal;
the orthonormal states are constructed by inverting the norm:
\begin{equation}
  |\widetilde{\Psi}^\alpha(W)\rangle
  = \sum_\beta{\left[\mathbf{1} +
           \mathbf{K}^2\right]^{-1/2}}_{\beta\alpha}
                                      |\Psi^\beta(W)\rangle
\label{orthonormalization}
\end{equation}
(see {\it e.g.\/} \cite{Newton}, eq.~(7.29)).
The state $\widetilde{\Psi}_B(M)$ in (\ref{PVB}) representing the 
intermediate $B$ baryon is constructed using 
(\ref{orthonormalization}), and is therefore normalized as
$
   \langle\widetilde{\Psi}_B(M)|\widetilde{\Psi}_B(M')\rangle
   = \delta(M-M')
$.
The construction of the orthonormal states corresponding
to intermediate isobars is discussed in Appendix~\ref{sec:wD}.

For a process in which the initial pion-nucleon system with invariant
mass $W$ decays into the (first) pion with momentum $\vec{k}$ and 
the pion-nucleon system with the quantum numbers of the intermediate 
baryon $B$, we write
\begin{equation}
   K_{\pi B\pi N}^{JT}(k,k_0,M) =  -\pi\sqrt{\omega_k E\over k\, W}
     \langle\Psi^N_{JT}(W)||V(k)||
                          \widetilde{\Psi}_B(M)\rangle\,.
\label{KCL-DN}
\end{equation}

The $\pi B$ to $\pi N$ transition matrix element of the multi-channel 
K matrix is
\begin{equation}
   K_{\pi N\pi B}^{JT}(k,k_1,M) =  -\pi\sqrt{\omega_k E\over k\,W}
     \langle\Psi^B_{JT}(W,M)||V(k)||\Psi_N\rangle\,.
\label{KCL-ND}
\end{equation}
The matrix element corresponding to the $\pi B$ to $\pi B'$ 
transition ({\it e.g.\/} 
$\pi(k_1)+\Delta(M)\rightarrow \pi(k)+\Delta(M')$) is given by
\begin{eqnarray}
   \kern-12pt K_{\pi B'\pi B}^{JT}(k,k_1,M',M) &&
\nonumber \\ && \kern-72pt
   =  -\pi\sqrt{\omega_k E\over k\, W}
   \langle\Psi^B_{JT}(W,M)||V(k)||\tilde{\Psi}_{B'}(M')\rangle\,.
\label{KCL-BB}
\end{eqnarray}

\subsection{\label{sec:sigma}Channels including the $\sigma$-meson}

We consider here the situation in which the decay proceeds
through an unstable meson and a baryon.
We derive the expression for the K matrix in the simplest
case of the $\sigma N$ channel (in the P11 wave)
and the $\Delta\pi$ channel (in the P33 wave) 
which dominate the inelastic processes in the 
energy range below $\sim 1700$~MeV for these two partial waves.
The extension of the method to other unstable mesons is
straightforward but more complicated because of a larger
number of involved channels.

The $\sigma$-meson appears as a chiral partner of the pion
in several versions of chiral quark models which also provide
the form of its coupling to the quark core, {\it e.g.\/} in the
linear $\sigma$-model with quarks \cite{Birse84,Ripka84,BB8586} 
or in various bosonized versions of the Nambu--Jona-Lasino model 
\cite{NJL:rev,ripka:book,BG03}.
In the non-linear realizations of the models its coupling
is realized by two correlated pions, 
see {\it e.g.\/} \cite{cooper86,kamano060504}.

In this work we use a purely phenomenological approach and 
take the form (\ref{Hpi}) for the $\sigma$-meson coupling.
We assume that only $s$-wave $\sigma$-mesons couple to the
quark core such that the interaction vertex takes the form:
$$
   \widetilde{V}_\mu(k) 
   = V^\mu(k)\,w_\sigma(\mu)    \,,
\qquad
    V^\mu(k) 
   = G_\sigma{k\over\sqrt{2\omega_{\mu k}}}\,.
$$
Apart from the momentum $k$, the one-$\sigma$-meson states are 
labeled by the invariant mass of the two-pion system, 
$2m_\pi<\mu<\infty$.
Here $\omega_{\mu k}^2 = k^2 + \mu^2$ 
and $w_\sigma(\mu)$ is a normalized mass distribution function 
centered around the nominal value of the $\sigma$-meson mass 
with the corresponding width, modeling the resonant decay into 
two pions.

In analogy with (\ref{KCL}), (\ref{KCL-DN}),  (\ref{KCL-ND}), 
and (\ref{KCL-BB}), we first introduce the matrix elements of 
the K matrix referring to the $\sigma N$ channel
in the case of the P11 partial waves. 
They involve the $\sigma$-meson with the energy and momentum 
$\omega_\mu$ and $k_\mu$ and the nucleon with the energy 
$E_N^\mu=\sqrt{M_N^2 +k_\mu^2}$ on the one side, and on the 
other side either the pion ($\omega_0$, $k_0$) and 
the nucleon, the pion ($\omega_1$, $k_1$) and the intermediate
baryon with the invariant mass $M$, or another $\sigma$-meson 
and the nucleon:
\begin{eqnarray}
K_{\sigma N \pi N}^{\h\h}(W,\mu) &=& -\pi\mathcal{N}_\mu\,
  \langle\Psi_{\h\h}^N(W)|{\widetilde{V}{}^\mu}(k_\mu)|\Psi_N\rangle\,,
\nonumber\\
K_{\pi N\sigma N}^{\h\h}(W,\mu) &=& -\pi\mathcal{N}_{0}\,
   \langle\Psi_{\h\h}^\sigma(W,\mu)||V(k_0)||\Psi_N\rangle\,, 
\nonumber\\
K_{\sigma N \pi B}^{\h\h}(W,\mu,M) &=& 
   -\pi\mathcal{N}_\mu\,
        \langle\Psi_{\h\h}^B(W,M)|
       {\widetilde{V}{}^\mu}( k_\mu)|\Psi_N\rangle\,,
\nonumber\\
K_{\pi B\sigma N}^{\h\h}(W,M,\mu) &=& -\pi\mathcal{N}_{1}\,
           \langle\Psi_{\h\h}^\sigma(W,\mu)||
                V(k_1)||\widetilde{\Psi}_B(M)\rangle\,,
\nonumber\\
K_{\sigma N\sigma N}^{\h\h}(W,\mu,\mu') &=& 
 -\pi\mathcal{N}_\mu\,
     \langle\Psi_{\h\h}^\sigma(W,\mu')|
          {\widetilde{V}{}^\mu}(k_\mu)|\Psi_N\rangle\,,
\nonumber\\
\label{Kmatrixsigma}
\end{eqnarray}
where $\mathcal{N}_0$ and $\mathcal{N}_1$ have been defined in 
(\ref{calN0}) and (\ref{calN1}),
$$
  \mathcal{N}_\mu = 
   \sqrt{\omega_\mu E_N^\mu(k_\mu)\over k_\mu W}\,,
$$
and
\begin{equation}
\omega_\mu = {W^2-M_N^2+\mu^2\over 2W}\,,
\qquad
E_N^\mu(k_\mu) = W - \omega_\mu\,.
\label{kinemasigma}
\end{equation}

In the P33 partial wave the decay 
into two correlated $l=0$ pions can proceed only through the 
intermediate $s$-wave $\sigma$-meson and the $\Delta$ isobar.
In this case the channel is labeled by the invariant mass of 
the two correlated pions, $\mu$, and the invariant mass of the 
pion-nucleon system from the $\Delta$ resonance, $M$.
The corresponding elements of the K matrix contain four invariant 
masses which is rather difficult to handle computationally. 
In sect.~\ref{sec:average} we discuss a method which simplifies 
the calculation by averaging over the $\Delta$  invariant mass 
such that the dependence on $M$ is eliminated.
The matrix elements then assume the same form as 
(\ref{Kmatrixsigma}) with $N$ replaced by $\Delta$, 
$M_N$ by $\bar{M}$, and $E_N^\mu$ by 
$\bar{E}^\mu=\sqrt{\bar{M}^2 +k_\mu^2}$.

In the P11 and P33 case the $\rho$-meson can be included in 
a similar way but since the experimental data \cite{PDG2004}
indicate that the contribution of the $\rho$-meson is almost 
negligible in the energy region under consideration we do not 
include it in our calculation.

\subsection{Constructing the T and the S matrix}

The multichannel K matrix acquires the following form
\begin{center}
\begin{tabular}{|l|l|l|}
\hline
$K_{NN}$            & $K_{NB}(M')$        & $K_{N\sigma}(\mu')$ \\
\hline
$K_{B N}(M)$   & $K_{BB'}(M,M')$ & $K_{B\sigma}(M,\mu')$ \\
\hline
$K_{\sigma N}(\mu)$ & $K_{\sigma B}(\mu,M)$ 
 & $K_{\sigma\sigma}(\mu,\mu')$ \\
\hline
\end{tabular}
\end{center}
where we have used a shorthand notation $B$ for the $\pi B$ channel 
and $\sigma$ for either the $\sigma N$ or the $\sigma\Delta$ channel.
The T matrix is related to the K matrix through (\ref{TSmatrix}).
In the inelastic channel the matrix elements depend on the 
continuous variable ($M$ or $\mu$), 
yielding a set of coupled integral (Heitler) equations:
\begin{eqnarray}
&&T_{NN} = K_{NN} +\i T_{NN}K_{NN}
\nonumber \\ && 
\kern12pt 
+\>\i\sum_B\int_{M_N+m_\pi}^{W-m_\pi}\d M\,T_{NB}(M) 
                            K_{B N}(M)
\nonumber \\ && 
\kern12pt 
+\>\i\int_{2m_\pi}^{W-M_N}\d\mu\,T_{N\sigma}(\mu) 
                            K_{\sigma N}(\mu)\,,
\nonumber\\
&&T_{N B}(M) = K_{N B}(M)  +\i T_{NN}K_{NB}(M)
\nonumber \\ &&
\kern12pt
+\>\i\sum_{B'}\int_{M_N+m_\pi}^{W-m_\pi}\d M'\,T_{NB'}(M') 
         K_{B'B}(M',M)
\nonumber \\ &&
\kern12pt
+\>\i\int_{2m_\pi}^{W-M_N}\d\mu\,T_{N\sigma}(\mu) 
         K_{\sigma B}(\mu,M)\,,
\nonumber\\
&&T_{N \sigma}(\mu) = K_{N \sigma}(\mu)  +\i T_{NN}K_{N\sigma}(\mu)
\nonumber \\ &&
\kern12pt
+\>\i\sum_{B}\int_{M_N+m_\pi}^{W-m_\pi}\d M\,T_{NB}(M) 
         K_{B\sigma}(M,\mu)
\nonumber \\ &&
\kern12pt
+\>\i\int_{2m_\pi}^{W-M_N}\d\mu'\,T_{N\sigma}(\mu') 
         K_{\sigma\sigma}(\mu',\mu)\,.
\label{eq4T}
\end{eqnarray}
In the P33 case the nucleon mass $M_N$ in the integral over
$\mu$ is replaced by the $W$-dependent averaged invariant mass
of the intermediate $\Delta$.

The unitarity of the S matrix it fulfilled provided the K matrix 
is real and symmetric which is especially important when we use 
approximate methods; in such a case it is considerably more 
advantageous to use a certain ansatz for the K matrix (or, 
equivalently for the principal-value state) rather than for 
the T matrix since in the latter case the unitarity has to be 
enforced at each step of the calculation.
Let us remark that for a general chiral quark model, the 
K~matrix and the corresponding principal-value state can be 
calculated variationally using the Kohn variational principle
\begin{equation}
   \langle\delta\Psi^\mathrm{P}|H-W|\Psi^\mathrm{P}\rangle = 0\,,
\label{Kohn}
\end{equation}
where  $\Psi^\mathrm{P}$ is a suitably chosen trial state.


\section{\label{sec:integral}The integral equations for the K matrix}

\subsection{Ansaetze for the channel PV states}

In the formal expressions for the PV states (\ref{PVN}) and 
(\ref{PVB}) the interaction $V(k)$ generates bare quark states
with quantum numbers different from the ground state, as well 
as superpositions of bare quark states and one or more mesons.
We choose a particular ansatz which implies the proper 
asymptotic behavior of different channels consisting of 
a meson and a baryon carrying its own meson cloud.
The form of the pion state in a $\pi B'$ channel can be 
read-off from the general relation (\ref{commute1})
holding for the eigenstate of the Hamiltonian (\ref{Hpi}).
Multiplying (\ref{commute1}) for $\Psi_A=\Psi_{JT}^{B}$ by 
$\langle\widetilde{\Psi}_{B'}(k)|$ we obtain
\begin{eqnarray}
&&  (\omega_k+E_{B'}(k)-W)\langle\widetilde{\Psi}_{B'}(k)|
  a_{mt}(k)|\Psi_{JT}^{B}(W)\rangle
\nonumber \\ && \hspace{60pt}
   = - \langle\widetilde{\Psi}_{B'}(k)|
            V_{mt}^\dagger(k)|\Psi_{JT}^{B}(W)\rangle\,,
\label{eq4chi0}
\end{eqnarray}
with $E_{B'}(k)=\sqrt{M'{}^2 + k^2}$, where $M'$ is either
the mass of a stable baryon or the invariant mass of the
$\pi N$ subsystem corresponding to the intermediate baryon $B'$.
For the elastic channel we assume the following ansatz: 
\begin{eqnarray}
|\Psi_{JT}^N(W)\rangle 
 &=& \mathcal{N}_0\,
\biggl\{
        [a^\dagger(k_0)|\Psi_N(k_0)\rangle]^{JT} 
   + \sum_\mathcal{R} c_\mathcal{R}^N(W)|\Phi_\mathcal{R}\rangle
\biggr. \nonumber \\ && \biggl. 
\kern -40pt
   + \int\d k\,{\chi_{JT}^{NN}(k,k_0)\over\omega_k+E_N(k)-W}\,
      [a^\dagger(k)|\Psi_N(k)\rangle]^{JT}
\biggr. \nonumber \\ && \biggl. 
\kern -40pt
   + \sum_{B}\int\d M\kern-3pt\int\d k\,
     {\chi_{JT}^{BN}(k,k_0,M)\over\omega_k+E_{B}(k)-W}
      [a^\dagger(k)|\widetilde{\Psi}_{B}(M)\rangle]^{JT}
\biggr. \nonumber \\ && \biggl. 
\kern -40pt
   + \int\d \mu \int\d k\,
       {\chi_{JT}^{\sigma N}(k,k_0,\mu)\over\omega_{\mu k}+E_{JT}(k)-W}\,
       b^\dagger(k)|\widetilde{\Psi}_{JT}\rangle
   \biggr\}\,.
\label{PsiN}
\end{eqnarray}
The first term, as discussed in the previous section, defines 
the channel and determines the normalization, the second term 
is the sum over bare quark states, denoted as $\Phi_\mathcal{R}$, 
with quantum numbers of the channel.
The amplitudes $\chi$ are proportional to the amplitudes 
(\ref{eq4chi0}); the first term corresponds to the one-pion state
on top of the ground state, the terms in the sum to the one-pion 
states around different excited states, and the last term to the 
one-$\sigma$ state around either the nucleon (P11) or the $\Delta$ 
(P33), with $E_{JT}(k)$ denoting the energy of the respective baryon.
Above the one- and  two-pion thresholds these amplitudes are related 
to the elastic and inelastic elements of the on-shell K matrix:
\begin{eqnarray}
K_{NN}(W)      &=& \pi\,\mathcal{N}_0^2\,\chi_{JT}^{NN}(k_0,k_0)\,,
\nonumber\\
K_{BN}(W,M) &=& \pi\,\mathcal{N}_0\mathcal{N}_1\,
             \chi_{JT}^{BN}(k_1,k_0,M)\,,
\nonumber\\
K_{\sigma N}(W,\mu) &=& \pi\,\mathcal{N}_0\mathcal{N}_\mu\,
             \chi_{JT}^{\sigma N}(k_\mu,k_0,\mu)\,.
\label{chi2KHN}
\end{eqnarray}
In the P11 case, one of the $\Phi_\mathcal{R}$ states is
the nucleon.

The ansatz can be simplified by realizing that the main 
contribution to the integrals over the invariant mass $M$, 
in (\ref{PsiN}), comes from $M$ close to $M_B$, {\it i.e.\/} 
the resonant energy of the isobar $B$.
In appendix~\ref{sec:wD} we show that in such a case one can write 
the state $\widetilde{\Psi}_B$ in a simplified form
\begin{equation}
 |\widetilde{\Psi}_B(M)\rangle
\equiv
   w_B(M)|\widehat{\Psi}_B(M)\rangle\,,
\label{tildePsiD}
\end{equation}
where $\widehat{\Psi}_B$ is dominated by the bare quark
configuration and only weakly depends on $M$.
The function $w_B(M)$ can be identified with the mass distribution 
function $\sigma(M)$ multiplied by the kinematic corrections 
-- the Blatt-Weiss\-kopf barrier-penetration factor -- that ensure 
proper threshold behavior (see {\it e.g.\/} \cite{manley92}).
Using this approximation the integration over 
$M$ in (\ref{PsiN}) selects $M=M_B$ and 
similarly $\mu=m_\sigma$ in the last term, yielding
\begin{eqnarray}
 |\Psi_{JT}^N(W)\rangle & = &\mathcal{N}_0\,
\biggl\{
    [a^\dagger(k_0)|\Psi_N(k_0)\rangle]^{JT} 
    +   \sum_\mathcal{R} c_\mathcal{R}^N(W)|\Phi_\mathcal{R}\rangle
\biggr. \nonumber \\ && \biggl. 
\kern -20pt
   + \int\d k\,{\chi_{JT}^{NN}(k,k_0)\over\omega_k+E_N(k)-W}\,
      [a^\dagger(k)|\Psi_N(k)\rangle]^{JT}
\biggr. \nonumber \\ && \biggl. 
\kern -20pt
   + \sum_B\int\d k\,
     {\widehat{\chi}_{JT}^{B N}(k,k_0,M_B)\over\omega_k+E_B(k)-W}
      [a^\dagger(k)|\widehat{\Psi}_B(M_B)\rangle]^{JT}
\biggr. \nonumber \\ && \biggl. 
\kern -20pt
   + \int\d k\,
       {\widehat{\chi}_{JT}^{\sigma N}(k,k_0,m_\sigma)\over
             \tilde{\omega}_k+E_{JT}(k)-W}\,
       b^\dagger(k)|\widehat{\Psi}_{JT}\rangle
   \biggr\}\,,
\label{PsiNa}
\end{eqnarray}
where $\tilde{\omega}_k = \sqrt{k^2+m_\sigma^2}$.
We have introduced
$$
\chi_{JT}^{BN} = w_B(M)\widehat{\chi}_{JT}^{BN}
\quad\hbox{and}\quad
\chi_{JT}^{\sigma N} = w_\sigma(\mu)\widehat{\chi}_{JT}^{\sigma N}\,.
$$

The inelastic channels corresponding to the $\pi B$ and 
$\sigma N$ channels can be written in compact forms
\begin{eqnarray}
&& \hskip-20pt|\Psi_{JT}^B(W,M)\rangle  
\nonumber \\
&=& \mathcal{N}_B\,w_B(M)\,
\biggl\{
     [a^\dagger(k_1)|\widehat{\Psi}_B(M)\rangle]^{JT} 
    + \sum_{\mathcal{R}} \widehat{c}_{\mathcal{R}}^B(W,M)
              |\Phi_{\mathcal{R}}\rangle
\biggr. \nonumber \\ && 
   + \int\d k\,
     {\widehat{\chi}_{JT}^{NB}(k,k_1,M)\over\omega_k+E_N(k)-W}\,
      [a^\dagger(k)|\Psi_N(k)\rangle]^{JT}
\nonumber \\ && 
   + \sum_{B'}\int\d k\,
       {\widehat{\chi}_{JT}^{B' B}(k,k_1,M_{B'},M)
                         \over\omega_k+E_{B'}(k)-W}\,
      [a^\dagger(k)|\widehat{\Psi}_{B'}(M_{B'})\rangle]^{JT}
\nonumber \\ && \biggl.
   + \int\d k\,
       {\widehat{\chi}_{JT}^{\sigma B}(k,k_1,m_\sigma,M)
                     \over\tilde{\omega}_k+E_{JT}(k)-W}\,
       a_\sigma^\dagger(k)|\widehat{\Psi}_{JT}\rangle
   \biggr\}
\label{PsiBa}
\end{eqnarray}
and
\begin{eqnarray}
&& \hskip-20pt|\Psi_{JT}^\sigma(W,\mu)\rangle  \nonumber \\
&=& \mathcal{N}_\sigma\,w_\sigma(\mu)\,
\biggl\{
     a_\sigma^\dagger(k_\mu)|\widehat{\Psi}_{JT}\rangle 
    + \sum_{\mathcal{R}} \widehat{c}_{\mathcal{R}}^\sigma(W,\mu)
                    |\Phi_{\mathcal{R}}\rangle
\biggr. \nonumber \\ && 
   + \int\d k\,
     {\widehat{\chi}_{JT}^{N\sigma}(k,k_\mu,\mu)\over\omega_k+E_N(k)-W}\,
      [a^\dagger(k)|\Psi_N(k)\rangle]^{JT}
\nonumber \\ && 
   + \sum_{B'}\int\d k\,
       {\widehat{\chi}_{JT}^{B'\sigma}(k,k_\mu,M_{B'},\mu)
                         \over\omega_k+E_{B'}(k)-W}\,
      [a^\dagger(k)|\widehat{\Psi}_{B'}(M_{B'})\rangle]^{JT}
\nonumber \\ && \biggl.
   + \int\d k\,
       {\widehat{\chi}_{JT}^{\sigma\sigma}(k,k_\mu,m_\sigma,\mu)
                     \over\tilde{\omega}_k+E_{JT}(k)-W}\,
       a_\sigma^\dagger(k)|\widehat{\Psi}_{JT}\rangle
   \biggr\}\,.
\label{Psisia}
\end{eqnarray}
Here we have assumed the following factorization
\begin{equation}
\chi_{JT}^{H'H} = w_H(m_H) w_{H'}(m_H')
                     \widehat{\chi}_{JT}^{H'H}\,,
\quad
c_{\mathcal{R}}^H = w_H(m_H)\widehat{c}_{\mathcal{R}}^H\,.
\label{factorization}
\end{equation}
where $H$ stands for either the $\pi B$ channels or $\sigma N$, 
and $m_H$ is the invariant mass (either $M$ or $\mu$).
Above the two-pion threshold the meson amplitudes are related to
the K matrix by
\begin{equation}
  K_{H'H}(W,m_H',m_H)   =
\pi\,\mathcal{N}_{H'}\mathcal{N}_H\,
             \chi_{JT}^{H'H}(k_{H'},k_H,m_H',m_H) \,.
\label{chi2K}
\end{equation}
The requirement that the K matrix be symmetric imposes the 
constraint
$
{\chi}_{JT}^{H'H}(k',k) = {\chi}_{JT}^{HH'}(k,k')
$
on the pion amplitudes.
The ansaetze therefore ensure that the scattering amplitudes 
are directly proportional to the on-shell K matrix needed in
the equation for the T matrix (\ref{eq4T}). 
For the off-shell matrices, equality (\ref{chi2K})
is not fulfilled in general.

\subsection{Derivation of the coupled integral equations}

The equations for the pion amplitudes $\chi$ and the
coefficients $c_\mathcal{R}$ in the ansaetze (\ref{PsiN}),
(\ref{PsiBa}) and (\ref{Psisia})
are derived either from (\ref{Kohn}) or directly from
the commutation relations (\ref{commute1}) which hold
for our particular choice of the quark-pion interaction.

By requiring stationarity with respect to the variation of 
the coefficients $c_\mathcal{R}^N$, $c_\mathcal{R}^B$ and  
$c_\mathcal{R}^\sigma$ we get
\begin{eqnarray}
&& \hspace{-40pt} (W-M_\mathcal{R}^0)c_\mathcal{R}^N(W) 
\nonumber\\ 
&=&  V_{N\mathcal{R}}(k_0) + \int\d k\,{V_{N\mathcal{R}}(k)\,
      \chi_{JT}^{NN}(k,k_0)
                 \over \omega_k+E_N(k)-W }
\nonumber\\ && 
+  \sum_{B'}\int\d k\,{ V_{B' \mathcal{R}}^{M_{B'}}(k)\,
       \widehat{\chi}_{JT}^{B' N}(k,k_0,M_{B'})
        \over \omega_k+E_{B'}(k)-W}
\nonumber\\ && 
+    \int\d k\,{ V_{N \mathcal{R}}^{m_\sigma}(k)\,
       \widehat{\chi}_{JT}^{\sigma N}(k,k_0,m_\sigma)
        \over \tilde{\omega}_k+ E_{JT}(k)-W}\,,
\label{eq4cRN}
\end{eqnarray}
\begin{eqnarray}
&&   \hspace{-40pt} 
(W-M_\mathcal{R}^0)\widehat{c}_{\mathcal{R}}^{B}(W,M)
\nonumber\\
&=&
 V_{B\mathcal{R}}^{M}(k_1) 
  + \int\d k\,{V_{N\mathcal{R}}(k)\,
     \widehat{\chi}_{JT}^{NB}(k,k_1,M)
          \over \omega_k+E_N(k)-W}
\nonumber\\  && 
   + \sum_{B'}\int\d k\,
       {V_{B' \mathcal{R}}^{M_{B'}}(k) 
        \widehat{\chi}_{JT}^{B'B}(k,k_1,M_{B'},M) 
        \over 
             \omega_k+ E_{B'}(k)-W}\,
\nonumber\\  && 
   +    \int\d k\,
       {V_{N \mathcal{R}}^{m_\sigma}(k) 
        \widehat{\chi}_{JT}^{\sigma B}(k,k_1,m_\sigma,M) 
        \over 
             \tilde{\omega}_k+ E_{JT}(k)-W}\,,
\label{eq4cRB}
\end{eqnarray}
\begin{eqnarray}
&& \hspace{-40pt} 
(W-M_\mathcal{R}^0)\widehat{c}_\mathcal{R}^\sigma(W,\mu) 
\nonumber\\ 
&=&  V^\mu_{\sigma\mathcal{R}}(k_\mu) 
   + \int\d k\,{V_{N \mathcal{R}}(k)\,
         \widehat{\chi}_{JT}^{N\sigma}(k,k_\mu,\mu)
                 \over \omega_k+E_N(k)-W}
\nonumber\\ && 
+  \sum_{B'}\int\d k\,{ V_{B' \mathcal{R}}^{M_{B'}}(k)\,
       \widehat{\chi}_{JT}^{B' \sigma}(k,k_\mu,M_{B'},\mu)
        \over \omega_k+E_{B'}(k)-W}
\nonumber\\ && 
+    \int\d k\,{ V_{N \mathcal{R}}^{m_\sigma}(k)\,
       \widehat{\chi}_{JT}^{\sigma \sigma}(k,k_\mu,m_\sigma,\mu)
        \over \tilde{\omega}_k+ E_{JT}(k)-W}\,.
\label{eq4cRs}
\end{eqnarray}
Here
\begin{eqnarray*}
  V_{N\mathcal{R}}(k) &=& \langle\Phi_\mathcal{R}||V(k)||\Psi_N\rangle
            = Z_N^{-1/2}\langle\Phi_\mathcal{R}||V(k)||\Phi_N\rangle\,,
\\
  V_{B\mathcal{R}}^{M}(k) &=& 
\langle\Phi_\mathcal{R}||V(k)||\widehat{\Psi}_B(M)\rangle
    = Z_B^{-1/2}\langle\Phi_\mathcal{R}||V(k)||\Phi_B\rangle\,,
\\
  V_{N\mathcal{R}}^{\mu}(k) &=& 
\langle\Phi_\mathcal{R}||V^\mu(k)||\widehat{\Psi}_N\rangle
    = Z_N^{-1/2}\langle\Phi_\mathcal{R}||V^\mu(k)||\Phi_N\rangle\,,
\end{eqnarray*}
where $Z_B$ is the wave-function normalization, while\break
$\langle\Phi_\mathcal{R}||V(k)||\Phi_N\rangle$ and
$\langle\Phi_\mathcal{R}||V(k)||\Phi_B\rangle$ 
are obtained from the underlying quark model.
In the P33 case $V_{N\mathcal{R}}^{\mu}(k)$ is replaced
by $V_{\Delta\mathcal{R}}^{\mu}(k)$.

In the P11 case, one of the states $\Phi_\mathcal{R}$ is replaced 
by the (exact) ground state in which the requirement of stationarity
is equivalent to the requirement that the channel states are
orthogonal to the ground state for $W>M_N$.
In this case the mass of the bare state $M_\mathcal{R}^0$ in 
(\ref{eq4cRN}), (\ref{eq4cRB}) and (\ref{eq4cRs}) is replaced by 
the ground-state mass $M_N$, while the matrix elements are given by
\begin{eqnarray}
 V_{N\mathcal{R}}(k)
&\rightarrow&
   (W-M_N){\langle\Psi_N||V(k)||\Psi_N\rangle\over 
           \omega_k + E_N(k) - W}\,,
\label{B2NN}
\\
 V_{B \mathcal{R}}^{M}(k)
&\rightarrow&
   (W-M_N){\langle\Psi_N||V(k)||\widehat{\Psi}_B(M)\rangle\over 
           \omega_k + E_B(k) - W}\,.
\label{B2ND}
\end{eqnarray}

Requiring stationarity with respect to pion amplitudes leads 
to the familiar  Lippmann-Schwinger equation for the K matrix.
The equation for the $\chi_{JT}^{NN}$ amplitude which is related 
to the elastic part of the K matrix is obtained from 
(\ref{eq4chi0}) for $B=B'=N$.
Using our ansatz (\ref{PsiNa}) and taking $M_J=M_T=\half$, 
we obtain, after multiplying (\ref{eq4chi0}) by 
$\CG{\h}{\h-m}{1}{m}{J}{\h}\CG{\h}{\h-t}{1}{t}{T}{\h}$
and summing over $m$ and $t$,
\begin{eqnarray}
\chi_{JT}^{NN}(k,k_0)
&=& {\cal K}^{NN}(k,k_0) 
    -\sum_\mathcal{R} c_\mathcal{R}^N(W) V_{N\mathcal{R}}(k)
\nonumber\\ && 
 + \int\d k'\,{{\cal K}^{NN}(k,k')\,\chi_{JT}^{NN}(k',k_0)\over
                    \omega_k'+E_N(k')-W}
\nonumber\\ && 
 + \sum_B \int\d k'\,{{\cal K}_{M_B}^{NB}(k,k')\,
                 \widehat{\chi}_{JT}^{B N}(k',k_0,M_B)
                    \over\omega_k'+E_B(k')-W}
\nonumber\\ && 
 + \int\d k'\,{{\cal K}_{m_\sigma}^{N\sigma}(k,k')\,
                 \widehat{\chi}_{JT}^{\sigma N}(k',k_0,m_\sigma)
                    \over\tilde{\omega}_k'+E_{JT}(k')-W}\,,
\label{eq4chiN} 
\end{eqnarray}
where we have introduced the kernels
\begin{eqnarray}
{\cal K}_M^{NB}(k,k')
 &=&
-\sum_{mtm't'}\langle\Psi_N(k)|
a^\dagger_{m't'}(k')
\nonumber \\ && 
\kern-30pt
\times 
\Bigl[V^\dagger_{mt}(k)
            +(\omega_k+E_N(k)-W)a_{mt}(k)\Bigr]|
      \widehat{\Psi}_B(M)\rangle
\nonumber\\ && 
\kern-30pt
\times   
      \CG{J_B}{\h-m'}{1}{m'}{J}{\h}\CG{T_B}{\h-t'}{1}{t'}{T}{\h}
      \CG{\h}{\h-m}{1}{m}{J}{\h}\CG{\h}{\h-t}{1}{t}{T}{\h}\,.
\label{calKNBdef}
\end{eqnarray}
For $B=N$, $\widehat{\Psi}_B(M)$ reduces to $\Psi_N$ and
$M$ to $M_N$.

For the general amplitude involving the $\pi B$ channels we use our 
ansatz (\ref{PsiBa}) which yields, after using (\ref{factorization}) 
and canceling $w_B(M)w_B(M')$ on both sides,
\begin{eqnarray}
&&  \hspace{-30pt}\widehat{\chi}_{JT}^{B'B}(k,k_1,M',M)\,
\nonumber \\
&=&
  {\cal K}_{M'M}^{B'B}(k,k_1)
 -\sum_\mathcal{R} \widehat{c}_\mathcal{R}^{B}(W,M) V_{B' \mathcal{R}}^{M'}(k)
\nonumber\\ &&
    + \sum_{B''}\int\d k'\,
  {{\cal K}_{M'M_{B''}}^{B'B''}(k,k')\,
  \widehat{\chi}_{JT}^{B''B}(k',k_1,M_{B''},M) 
         \over \omega_k'+E_{B''}(k')-W}\,
\nonumber\\ &&
    + \int\d k'\,
  {{\cal K}_{M'm_\sigma}^{B'\sigma}(k,k')\,
  \widehat{\chi}_{JT}^{\sigma B}(k',k_1,m_\sigma,M) 
         \over \tilde{\omega}_k'+E_{JT}(k')-W}\,.
\label{eq4chiBB}
\end{eqnarray}
Here 
\begin{eqnarray*}
\kern-12pt {\cal K}_{MM'}^{BB'}(k,k')
&=&
   -\sum_{mtm't'}\langle\widehat{\Psi}_{B}(M)|
       a^\dagger_{m't'}(k')
\nonumber \\ && 
\kern-54pt
\times
\left[V^\dagger_{mt}(k)
            +(\omega_k+E(k)-W)a_{mt}(k)\right]
                     |\widehat{\Psi}_{B'}(M')\rangle
\nonumber\\ && 
\kern-54pt
\times
     \CG{J_B'}{\h-m'}{1}{m'}{J}{\h}\CG{T_B'}{\h-t'}{1}{t'}{T}{\h}
     \CG{J_B}{\h-m}{1}{m}{J}{\h}\CG{T_B}{\h-t}{1}{t}{T}{\h}\,.
\end{eqnarray*}
The form of (\ref{eq4chiBB}) justifies the factorization 
(\ref{factorization}) for the amplitude $\chi_{JT}^{B'B}$.
(The expression for ${\cal K}_{M'm_\sigma}^{B'\sigma}$
used in our calculation is given in sect.~\ref{sec:Born}.)

Equations (\ref{eq4chiN}) and (\ref{eq4chiBB}) imply the 
following form for the pion $\chi$ amplitudes:
\begin{eqnarray}
   \chi_{JT}^{NN}(k,k_0) &=& -\sum_\mathcal{R} c_\mathcal{R}^N(W)
                            \mathcal{V}_{N\mathcal{R}}(k) 
                               +\mathcal{D}^{NN}(k,k_0)\,,
\nonumber\\
\label{cal2chiN}\\
   \widehat{\chi}_{JT}^{B'B}(k,k_1,M',M) &=& 
      -\sum_\mathcal{R}\widehat{c}_\mathcal{R}^B(W,M)
        \mathcal{V}^{M'}_{B' \mathcal{R}}(k) 
\nonumber\\ && 
               +\mathcal{D}_{M'M}^{B'B}(k,k_1)\,,
\label{cal2chiBB}
\end{eqnarray}
where $\cal{V}$ are the dressed vertices
and $\cal{D}$ are the background parts of the amplitudes.

The amplitudes involving the $\sigma$-meson fulfill the same 
type of integral equations as the pion amplitudes with 
the kernels given in Appendix~\ref{kernels}. 
They assume the forms:
\begin{eqnarray}
   \widehat{\chi}_{JT}^{B \sigma}(k,k_\mu,M,\mu) &=& 
        -\sum_\mathcal{R}\widehat{c}_\mathcal{R}^\sigma(W,\mu)
         \mathcal{V}^{M}_{B \mathcal{R}}(k) 
               +\mathcal{D}_{M\mu}^{B\sigma}(k,k_\mu)\,,
\nonumber\\
\label{cal2chiBs}\\
   \widehat{\chi}_{JT}^{\sigma B}(k,k_1,\mu,M) &=& 
        -\sum_\mathcal{R} \widehat{c}_\mathcal{R}^B(W,M)
         \mathcal{V}^{\mu}_{\sigma \mathcal{R}}(k) 
               +\mathcal{D}_{\mu M}^{\sigma B}(k,k_1)\,,
\nonumber\\
\label{cal2chisB}
\\
   \widehat{\chi}_{JT}^{\sigma\sigma}(k,k_\mu,\mu',\mu) &=& 
        -\sum_\mathcal{R} \widehat{c}_\mathcal{R}^\sigma(W,\mu)
              \mathcal{V}^{\mu'}_{\sigma \mathcal{R}}(k) 
               +\mathcal{D}_{\mu'\mu}^{\sigma\sigma}(k,k_\mu)\,.
\nonumber\\
\label{cal2chiss}
\end{eqnarray}

\subsection{\label{sec:Born}The Born approximation for the K matrix}

The Born approximation for the K matrix consists in neglecting 
the terms in  (\ref{eq4cRN}), (\ref{eq4cRB}), (\ref{eq4cRs}), 
(\ref{eq4chiN}),  and (\ref{eq4chiBB}) involving the integrals.
The K matrix is then constructed from the meson amplitudes
(\ref{cal2chiN})--(\ref{cal2chiss}) using (\ref{chi2KHN})
and (\ref{chi2K}), and substituting the dressed vertices 
$\mathcal{V}_{B\mathcal{R}}$ by the corresponding bare vertices 
$V_{B\mathcal{R}}$ as well as $\mathcal{D}$ by $\mathcal{K}$.
The expressions for $\mathcal{K}^{HH'}$ are derived in 
Appendix~\ref{kernels}; note that they involve only the 
on-shell amplitudes which are not influenced by the 
approximation of separable kernels.
They acquire the form
$$
\mathcal{K}^{BB'}_{MM'}(k_1,k_1') = 
    \sum_{B''}   f_{BB'}^{B''}\,
         {2M_{B''}\,V_{B''B}^M(k_1')V_{B''B'}^{M'}(k_1)\over
        2E\omega_1' +  M_{B''}^2 - M^2 - m_\pi^2}
$$
and (in the P11 case)
\begin{eqnarray}
\mathcal{K}^{B\sigma}_{M\mu}(k_1,k_\mu) &=&
    \sum_{B'}
   {2M_B\,V^\mu_{B'B}(k_\mu)V_{B'N}(k_1)
               \over 2E\omega_\mu + M_{B'}^2 - M^2 - \mu^2}
\nonumber\\
    &=& \mathcal{K}^{\sigma B}_{\mu M}(k_\mu,k_1)\,,
\nonumber\\
\mathcal{K}^{\sigma\sigma}_{\mu\mu'}(k_{\mu},k_{\mu}') &=&  
    \sum_{B}
    {2M_B\,V^\mu_{BN}(k_\mu')V^\mu_{BN}(k_\mu)
      \over 2{E_N^\mu}'\omega_\mu+M_B^2-M_N^2-\mu^2}\,.
\nonumber\\
\label{KsigmaBorn}
\end{eqnarray}
The symmetry in the background terms follows from the 
symmetry of the denominator in the $u$-channel, {\it e.g.\/}\break
$2E_N\omega_\mu-M_N^2-\mu^2=2E_N^\mu\omega_0-M_N^2- m_\pi^2$.
Because we deal with $s$-wave scattering, the direct term
({\it i.e.\/} the first term in (\ref{cal2chiBs})) referring to 
the nucleon pole and the background part for $B=N$ almost 
cancel and can be dropped from the above sums.

In the P33 case the decay into two correlated $l=0$ pions 
proceeds through the $\sigma\Delta$ channel.
As discussed in sect.~\ref{sec:sigma} it is sensible to average
over the $\Delta$ invariant mass such that the matrix elements
depend only on the invariant mass of the two-pion system.
The averaging is discussed in the following.

\subsection{\label{sec:average}Averaging over invariant masses}

The averaging over the  $\Delta$ invariant mass in the 
$\sigma\Delta$ channel implies that the matrix elements
in the kernels  assume the same form as (\ref{KsigmaBorn})
with $N$ replaced by $\Delta$, and $V^\mu_{NB}(k)$ by
the averaged interaction matrix element defined as
\begin{equation}
   \bar{V}^\mu_{\Delta B}(k_\mu)^2 
= 
         {\bar{k}_\mu W\over\bar{\omega}_\mu\bar{E}^\mu}
          \int_{M_N+m_\pi}^{W-m_\pi}\kern-3pt
           \d M\,w_\Delta(M)^2\,\,
         \mathcal{N}_\mu^2\,V^\mu_{\Delta B}(k_\mu)^2,
\label{averageVmu}
\end{equation}
while the denominator in $\mathcal{K}^{\sigma\sigma}$ contains
$\bar{\omega}_\mu$, $\bar{k}_\mu$ and $\bar{E}^\mu$ evaluated
by (\ref{kinemasigma}) in which $M_N$ is replaced by 
the averaged invariant mass $\bar{M}$.

The averaged invariant masses $\bar{M}$ and $\bar{\mu}$
are found by suitable smooth numerical approximations 
approaching the nominal hadron masses for large $W$, 
while remaining close to either $M_N+m_\pi$ or $2m_\pi$,
for $W$ slightly above the two pion threshold.

The approximation of averaging over the invariant mass as
in (\ref{averageVmu}) can be applied to other matrix elements,
as well as to the K matrix and the T matrix themselves.
For the decay of a resonance $B'$ into a pion and an unstable
isobar $B$ which in turn decays into the nucleon and the
second pion, we introduce
\begin{equation}
   \bar{V}_{BB'}(\bar{k}_{1})^2 = 
     {\bar{k}_{1}W\over\bar{\omega}_{1}\bar{E}}
      \int_{M_N+m_\pi}^{W-m_\pi}
         \kern-6pt\d M\,w_B(M)^2
             \,\mathcal{N}_1^2\, V^M_{BB'}(k_1)^2,
\label{VBBaverage}
\end{equation}
where $\bar{\omega}_{1}$ and $\bar{E}$ are those 
of (\ref{kinema1}) evaluated at $M=\bar{M}$.

Similarly, for the decay of a resonance $B'$ through a baryon $B$
and a $\sigma$-meson which in turn decays into two pions, 
the matrix element averaged over the meson invariant mass reads
$$
   \bar{V}_{BB'}(\bar{k})^2 = 
     {\bar{k}W\over\bar{\omega}\bar{E}}
     \int_{2m_\pi}^{W-M_B}\d \mu\,
         w_\sigma(\mu)^2\,\,
         \mathcal{N}_\mu^2\, V_{BB'}^\mu(k_\mu)^2\,,
$$
where $\bar{\omega} = (W^2-M_B^2+\bar{\mu}^2)/2W$,
$\bar{E} = W - \bar{\omega}$ and 
$\bar{k}^2= \bar{\omega}^2+\bar{\mu}^2$.
The averaging procedure turns the integral Heitler equation 
(\ref{eq4T}) into a set of algebraic equations. 
Such an approximation does not influence the elastic channel since
the matrix elements involving unstable hadrons in this channel 
always appear only under the integral; 
in inelastic channels this means that the 
$M$-dependent amplitudes are replaced by some averaged value.
Identical averaging of amplitudes (\ref{VBBaverage}) is used in 
phenomenological analyses of $\pi N\rightarrow \pi\pi N$
reactions proceeding through the unstable intermediate hadron  
(see {\it e.g.\/} \cite{manley92}).

\subsection{\label{sec:solving}Solving the integral equations 
in the approximation of separable kernels}

In this section we solve the set of coupled integral equations 
(\ref{eq4cRN}), (\ref{eq4cRB}), (\ref{eq4chiN}), (\ref{eq4chiBB}), 
beyond the Born approximation for the K matrix.
The solution for the vertices yields a considerable enhancement
with respect to their bare values while the solution for the
coefficients $c_{\mathcal{R}}$ involves a considerable mixing of different
'bare' resonances (denoted by $\mathcal{R}$) appearing in our ansatz.
The method yields simultaneously the position of the resonance
as well as the pertinent wave-function renormalization.
Since the quark-$\sigma$ vertex is not as well determined 
as the quark-$\pi$ vertex, we treat the $\sigma$-meson vertices 
only in the Born approximation discussed above.

Inserting the ansaetze (\ref{cal2chiN})--(\ref{cal2chiss})
into the set of coupled equations we obtain
\begin{eqnarray}
\mathcal{V}_{N\mathcal{R}}(k) &=& V_{N\mathcal{R}}(k) + 
  \int\d k'\,{\mathcal{K}^{NN}(k,k')\mathcal{V}_{N\mathcal{R}}(k')
                        \over \omega_k'+E_N(k')-W}  
\nonumber\\ && + 
  \sum_{B'}\int\d k'\,{\mathcal{K}_{M_{B'}}^{NB'}(k,k')
                       \mathcal{V}^{M_{B'}}_{B'\mathcal{R}}(k')
                        \over \omega_k'+E_{B'}(k')-W}\,,
\label{eq4calVBN} \\ 
\mathcal{V}^{M}_{B\mathcal{R}}(k) &=& V_{B\mathcal{R}}^{M}(k) + 
  \int\d k'\, {\mathcal{K}_{M}^{BN}(k,k')\mathcal{V}_{N\mathcal{R}}(k')
                        \over \omega_k'+E_N(k')-W} 
\nonumber\\ && + 
 \sum_{B'}\int\d k'\, {\mathcal{K}_{MM_{B'}}^{BB'}(k,k')
             \mathcal{V}^{M_{B'}}_{B'\mathcal{R}}(k')
        \over  \omega_k'+E_{B'}(k')-W}\,.
\label{eq4calVBD} 
\end{eqnarray}
The background parts $\mathcal{D}^{NN}(k,k_0)$ and
$\mathcal{D}_M^{NB}(k,k_1)$ obey the integral equations 
of the type (\ref{eq4calVBN}) with the non-homo\-geneous terms 
$\mathcal{K}^{NN}(k,k_0)$ and $\mathcal{K}_M^{NB}(k,k_1)$,
respectively, while the $\mathcal{D}_M^{B N}(k,k_0)$ and
$\mathcal{D}_{M'M}^{B'B}(k,k_1)$ satisfy the integral equations 
of the type (\ref{eq4calVBD}) with the non-homogeneous terms 
$\mathcal{K}_M^{B N}(k,k_0)$ and 
$\mathcal{K}_{M'M}^{B'B}(k,k_1)$, respectively.

In appendix \ref{kernels} we introduce several approximations
which enable us to write the kernels in separable form:
\begin{eqnarray}
\mathcal{K}^{NN}(k,k') &=& 
     \sum_{B'} f_{NN}^{B'} {M_{B'}\over E_N}\,
  (\omega_0+\varepsilon_{B'}^N)
\nonumber \\ && \times
  {\mathcal{V}_{B'N}(k')\,\mathcal{V}_{B'N}(k)
  \over 
  (\omega_k'+\varepsilon_{B'}^N)(\omega_k+\varepsilon_{B'}^N)}\,,
\label{calKNN} \\
{\cal K}_M^{NB}(k,k') &=&   
   \sum_{B'} f_{NB}^{B'}\,{M_{B'}\over E}\,
   (\omega_1 + \varepsilon_{B'}^N)
\nonumber \\ && \times
  {\mathcal{V}_{B'N}(k')\,\mathcal{V}_{B'B}^M(k)
  \over 
  (\omega_k'+\varepsilon_{B'}^N)(\omega_k+\varepsilon_{B'}^B(M))}
\nonumber \\
&=& {\cal K}_{M}^{B N}(k',k)\,,
\\
{\cal K}_{MM'}^{BB'}(k,k')
&=& 
 \sum_{B''} f_{B'B}^{B''}\,{M_{B''}\over E'}\,
   (\omega_1'+\varepsilon_{B''}^B(M))
\nonumber \\ && \times
  {\mathcal{V}_{B''B}^M(k') \,\mathcal{V}_{B''B'}^{M'}(k)
       \over
            (\omega_k'+\varepsilon_{B''}^B(M))
            (\omega_k +\varepsilon_{B''}^{B'}(M'))}   \,,\quad
\label{calKDD}
\end{eqnarray}
where
$
   \varepsilon_{B'}^N = (M_{B'}^2 - M_N^2 - m_\pi^2)/2E_N
$
and
$
    \varepsilon_{B'}^B(M) = (M_{B'}^2 - M^2 - m_\pi^2)/2E
$,
while $f_{AB}^C$ are given by (\ref{fabc}).
Here $M_{B'}$ stands for the nominal (fixed) mass of the 
isobar $B'$, while the invariant mass $M$ pertinent 
to isobar $B$ is a variable.
Using the  separable kernels we are able to solve the system exactly.
This is important from the numerical point of view since it is
now possible to control the principal value integration over the 
poles of the kernel and thus avoid possible numerical instabilities.

For the coefficients $c^H_\mathcal{R}$ (here $H$ denotes $\pi N$, $\pi B$, 
$\sigma B$ channels) 
we obtain a set of algebraic equations
\begin{equation}
\sum_{\mathcal{R}'} A_{\mathcal{R}\mathcal{R}'}(W)c^H_{\mathcal{R}'}(W,m_H) 
   = b_\mathcal{R}^H(m_H)\,,
\label{eq4cH}
\end{equation}
where
\begin{eqnarray}
A_{\mathcal{R}\mathcal{R}'} &=&
 (W-M_\mathcal{R}^0)\delta_{\mathcal{R}\mathcal{R}'}
      +   \sum_{B'}
          \int\d k\,{\mathcal{V}^{M_{B'}}_{B'\mathcal{R}}(k)
                              V_{B'\mathcal{R}'}^{M_{B'}}(k)
                        \over \omega_k+E_{B'}(k)-W}\,,
\nonumber \\
b_\mathcal{R}^{B} &=&   V_{B\mathcal{R}}^M(k_1)
       + \sum_{B'} 
           \int\d k\,{\mathcal{D}_{MM_{B'}}^{B'B}(k,k_1)
                       V_{B'\mathcal{R}}^{M_{B'}}(k)
                        \over \omega_k+E_{B'}(k)-W}
\nonumber\\
         &=& \mathcal{V}_{B\mathcal{R}}^M(k_1)\,,
\nonumber\\
b_\mathcal{R}^\sigma &=&   V_{N\mathcal{R}}^\mu(k_\mu)\,.
\label{eq4cDa}
\end{eqnarray}
Here the sum over $B'$ includes also the ground state,
in the P33 case $V_{\sigma\mathcal{R}}$ is replaced
by $V_{\Delta\mathcal{R}}$.
The equalities between $b$ and $\mathcal{V}$ can be proved by
iterating equations for  $\mathcal{D}$ and $\mathcal{V}$.

Using the vector notation
$\vec{c}^H\equiv [c_\mathcal{R}^H, c_{\mathcal{R}'}^H, \ldots]^T$
and 
$\vec{\mathcal{V}}_H\equiv[\mathcal{V}_{H\mathcal{R}}, 
\mathcal{V}_{H\mathcal{R}'}, \ldots]^T$, 
the solution of (\ref{eq4cH}) can be  written in the form
$
           \vec{c}^H = {\bf A}^{-1}\vec{\mathcal{V}}_H\,.
$
The zeros of {\bf A} occur at the positions of the poles of 
the K matrix related to the resonance $\mathcal{R}$; 
we denote the corresponding energies by $M_\mathcal{R}$.
The procedure to determine the coefficients $c_\mathcal{R}$ is then
the following:
we first determine the zeros of the {\bf A}-matrix determinant; 
by adjusting the energies of the bare states, $M_\mathcal{R}^0$, 
we can 
force the poles of the K matrix to acquire some desired values.
(Note that in the case of several channels and strong background 
they do not coincide with the corresponding experimental values.)
Diagonalizing the {\bf A} matrix,
${\bf U}{\bf A}{\bf U}^T = {\bf D}$,
we write
\begin{eqnarray}
    {\bf D} &=& 
  \hbox{diag}[\lambda_\mathcal{R}, \lambda_{\mathcal{R}'}, \ldots]
\nonumber \\
  &\equiv& \hbox{diag}[Z_\mathcal{R}(W)(W-M_\mathcal{R}), 
                       Z_{\mathcal{R}'}(W)(W-M_{\mathcal{R}'}), 
\ldots]\,,\nonumber \\ &&
\label{defZ}
\end{eqnarray}
which defines the wave-function normalization $Z_\mathcal{R}$ 
pertinent to the resonance $\mathcal{R}$.
The solution can now be cast in the form
$$
 \vec{c}^H = {\bf U}^T{\bf D}^{-1}{\bf U}\vec{\mathcal{V}}_H\,.
$$
Finally, the resonant part of the $\chi$ amplitudes appearing in 
the expression for the K matrix ({\it e.g.\/} (\ref{cal2chiN})) 
takes the form
\begin{eqnarray}
   \chi^{H'H} &=&  -\vec{\mathcal{V}}_{H'}^T \vec{c}^H      
              =   - \vec{\mathcal{V}}_{H'}^T 
                {\bf U}^T{\bf D}^{-1}{\bf U}\vec{\mathcal{V}}_H
\nonumber \\
  &=& 
   - \sum_\mathcal{R} \widetilde{\cal V}_{H\mathcal{R}}\,
            {1\over Z_\mathcal{R}(W) (W-M_\mathcal{R})}\,
                    \widetilde{\cal V}_{H'\mathcal{R}}
\nonumber \\
  &=& 
   -\sum_\mathcal{R} \widetilde{c}^{H}_\mathcal{R}
                   \widetilde{\cal V}_{H'\mathcal{R}}\,,
\label{sol4chi}
\end{eqnarray}
where
\begin{equation}
    \widetilde{\cal V}_{H\mathcal{R}} = 
        \sum_{\mathcal{R}'}u_{\mathcal{R}\mathcal{R}'}{\cal V}_{H\mathcal{R}'}\,,
\qquad
    \widetilde{c}_\mathcal{R}^H = {\widetilde{\cal V}_{H\mathcal{R}}
             \over Z_\mathcal{R}(W) (W-M_\mathcal{R})}\,.
\label{calVmix}
\end{equation}
The interpretation of (\ref{calVmix}) is that the resonant states 
$\mathcal{R}$, $\mathcal{R}'$, $\dots$  
are not eigenstates of $H$ and therefore mix:
$$
   \widetilde{\Phi}_\mathcal{R} = \sum_{\mathcal{R}'} 
   u_{\mathcal{R}\mathcal{R}'} \Phi_{\mathcal{R}'}\,.
$$
The exception is the ground state -- which by assumption is the
eigenstate of $H$ -- for which the mixing of other resonances
because of (\ref{B2NN}) and (\ref{B2ND})
vanishes at the nucleon pole ($W=M_N$) 
and does not affect the $\pi NN$ coupling constant.

Note that neglecting the off-diagonal terms $A_{\mathcal{R}\mathcal{R}'}$ 
in (\ref{eq4cDa}) the expression for $Z_\mathcal{R}$ in the vicinity 
of the resonance assumes the familiar form
$$
    Z_\mathcal{R}(M_\mathcal{R}) 
    = 1 - \left.{\d\over\d W}\Sigma_\mathcal{R}(W)\right|_{W=M_\mathcal{R}}\,.
$$


\section{\label{sec:results}Results}

\subsection{\label{CBM}The parameters of the Cloudy Bag Model}

We analyze the capability of our approach in the framework 
of the Cloudy Bag Model (CBM) as one of the most popular examples 
of quark-pion dynamics.
The Hamiltonian of the model has the form (\ref{Hpi}) and 
(\ref{Vmt}) with
$$
  v(k) = {1\over2f}\,{k^2\over\sqrt{12\pi^2\omega_k}}\,
    {\omega^0_\mathrm{MIT}\over\omega^0_\mathrm{MIT}-1}\,
            {j_1(kR)\over  kR}\,,
$$
assuming three quarks in the $1s$ state.
The parameter $f$ corresponds to the pion decay constant $f_\pi$,
and $\omega^0_\mathrm{MIT}=2.043$.
It is a known drawback of the model that the $\pi NN$ coupling 
constant is underestimated, irrespectively of the bag radius, 
if $f$ is fixed to the experimental value $f_\pi=93$~MeV.
We therefore adopt  the conventional smaller value of $f=76$~MeV 
which reproduces 
the $\pi NN$ coupling constant.
The free parameters of the model are the bag radius $R$
and the energies of the bare quark states corresponding to
the nucleon and the excited states. 
We have also considered alternative forms of the $k$-dependence 
which avoid the typical oscillations due to the sharp cut-off at the 
bag surface but have found almost no change in the final results.
We use the same bag radius for the excited states as for the 
ground state; as a consequence, the matrix elements of the 
quark-pion interaction between the quark configurations with 
different spatial structure are all proportional to $v(k)$:
\begin{eqnarray}
   \kern-12pt\langle\Phi_{B'}||V(k)||\Phi_{B}\rangle &&
\nonumber \\ && \kern-48pt 
=r_{BB'}\,r_q\,v(k)\,
\langle J_{B'},T_{B'}||
    \sum_{i=1}^3 \sigma^i\tau^i||J_B,T_B\rangle\,,
\label{BVB}
\end{eqnarray}
where $r_q=1$ if both $B$ and $B'$ are in the $(1s)^3$ configuration,
$r_q=r_\omega$ for the transition between the  $(1s)^2(2s)^1$ 
configuration and the ground state, and $r_q={2\over3}+r_\omega^2$ 
if both $B$ and $B'$ are in the $(1s)^2(2s)^1$ configuration.
Here 
$$
r_\omega = 
{1\over\sqrt{3}}\,
  \biggl[\, {\omega_\mathrm{MIT}^1(\omega_\mathrm{MIT}^0-1)\over
             \omega_\mathrm{MIT}^0(\omega_\mathrm{MIT}^1-1)}\,
  \biggr]^{1/2}\,,
$$
with $\omega^1_\mathrm{MIT}=5.396$.
The parameter $r_{BB'}$ in (\ref{BVB}) allows us to tune the
chosen coupling constant relative to its $SU(6)$ value.
The value $r_{NN}=1$ is fixed by our choice of $f$.
We assume $r_{BB'}=r_{B'B}$.

In the P11 case the sum over $\mathcal{R}$ in (\ref{PsiNa}), 
(\ref{PsiBa}) and (\ref{Psisia}) includes beside the nucleon, 
the Roper $N(1440)$ and the $N(1710)$, and in the P33 case, 
the $\Delta(1232)$, $\Delta(1600)$ and $\Delta(1920)$.
We do not include further intermediate states and channels since
our present goal is to find a pattern common to the low-lying
Roper-like resonances.  This limits the validity of our approach
to energies below $\sim 1700\;\mathrm{MeV}$.
The approach can be extended in a straightforward way by including 
higher intermediate states as well as other channels.

\setlength{\tabcolsep}{10pt}
\begin{table}[ht]
\caption{The model parameters used in sect.~\ref{sec:results-Born}
(Born) and sect.~\ref{sec:results-full} (Full) for the 
P11 and P33 partial waves.
$M_B$ are the positions of the K-matrix poles,
$M_\RS$ corresponds to $N(1710)$.
The parameter $r_{BB'}$  is defined in (\ref{BVB})
 and
$\bar{r}_{\pi BB'}$ refers to the values used in the
kernels in the full calculation.}
\begin{center}
\begin{tabular}{lcccc}
\hline

\hline
     & \multicolumn{2}{c} {Born} & \multicolumn{2}{c} {Full} \\
\hline
 Parameter & P11 & P33               & P11 & P33  \\
\hline
$R$         & \multicolumn{4}{c} {0.83~fm}        \\
$M_\Delta$   & \multicolumn{4}{c} {1232~MeV}      \\
$M_R$        & \multicolumn{4}{c} {1520~MeV}      \\
$M_\DS$      & \multicolumn{4}{c} {1780~MeV}      \\
$M_\RS$      & \multicolumn{4}{c} {1870~MeV}      \\
\hline
$m_\sigma$        & \multicolumn{4}{c} {450~MeV}  \\
$\Gamma_\sigma$   & \multicolumn{4}{c} {550~MeV}  \\
$G_\sigma$        & \multicolumn{2}{c} {0.96} 
                  & \multicolumn{2}{c} {0.99}     \\
\hline
  $r_{N R}$          & 1.68 & 0.75 & \multicolumn{2}{c} {1.00}  \\
  $r_{N\Delta}$  & 0.83 & 1.39 & \multicolumn{2}{c} {1.00}  \\
  $r_{\Delta R}$ & \multicolumn{2}{c}{0.80} 
                               & \multicolumn{2}{c} {1.30}  \\
\hline
     $\bar{r}_{\pi N\Delta}$      &  &  & 1.30 & 1.12 \\
     $\bar{r}_{\pi \Delta\Delta}$ &  &  & 1.00 & 1.25 \\
     $\bar{r}_{\pi NN}$           &  &  & 1.00 & 1.00 \\
\hline

\hline
\end{tabular}
\end{center}
\label{tab:parameters}
\end{table}

In sect.~\ref{sec:results-Born} we first discuss the results 
of the Born approximation for the K matrix, and in 
sect.~\ref{sec:results-full} the results when the
integral equation for the K matrix is solved.
The parameters  in the two cases are displayed in
Table~\ref{tab:parameters}.
The parameters $M_\Delta$, $m_\sigma$ and $\Gamma_\sigma$ 
are kept fixed: $M_\Delta$ at the experimental position 
of the pole of the K matrix, while
from the recent analysis of Leutwyler \cite{leutwyler07}
we take $m_\sigma=450$~MeV and $\Gamma_\sigma=550$~MeV.
The values for $M_R$, $M_\DS$ and $M_\RS$, and for $G_\sigma$ 
are free  in the Born approximation as well as in the full
calculation.
The parameters $r_{BB'}$ defined in (\ref{BVB}) are free 
in the Born approximation; in the full calculation they 
are kept at the values predicted by the quark model except 
for the value of the $\pi\Delta R$ coupling.
The $\bar{r}_{\pi BB}$ correspond to the averaged values 
of the dressed vertices and are explained 
in sect.~\ref{sec:results-full}.

\subsection{\label{sec:results-Born} The Born approximation}

We first analyze the P33 partial amplitudes.
The results turn out to be almost insensitive to the value of the 
bag radius so the only parameters to adjust are the positions of 
the resonances $M_\Delta$ and $M_\DS$ and the relative coupling 
strengths $r_{N\Delta}$ and $r_{N \DS}\equiv r_{N\Delta}r_{N R}$.
(Note that our value for $M_\DS$ is the position of the K-matrix 
pole and should not be identified with the nominal value
of the resonance invariant mass.)

\begin{figure}[h!]
\includegraphics[width=8cm]{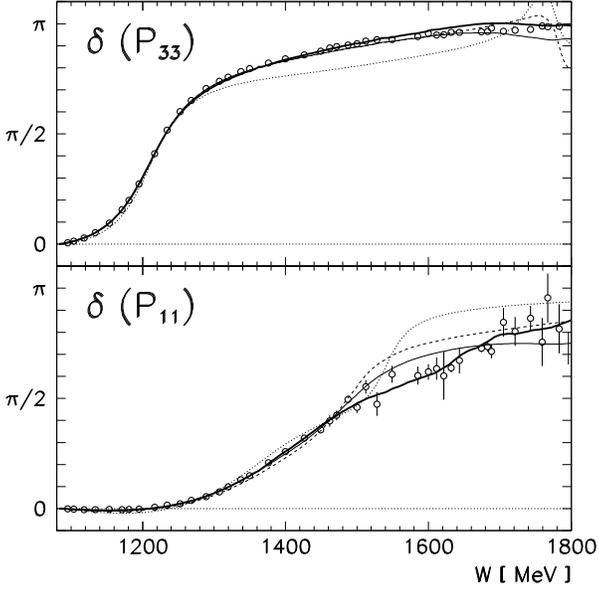}
\caption{The P33 (top panel) and P11 (bottom panel) phase shifts 
in various approximations.  
Born approximation: $\pi N$ and $\pi\Delta$ channels, resonant 
terms only (dotted lines), with background (dashed lines), 
adding the $\sigma$ channel (thin solid lines).
Full calculation: thick solid lines.
The data points in this and subsequent figures are from 
the SAID $\pi N\to\pi N$ partial-wave analysis 
\protect\cite{SAID06} unless noted otherwise.
The model parameters are given in Table~\ref{tab:parameters}.}
\label{fig:delta3311}
\end{figure}

\begin{figure}[h!]
\includegraphics[width=8cm]{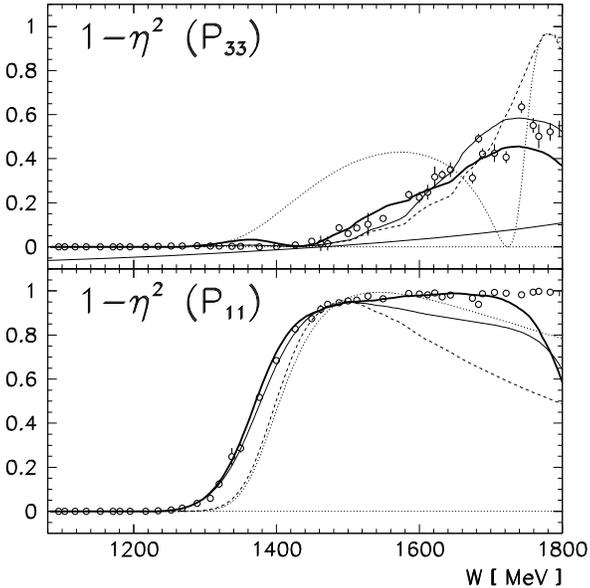}
\caption{Inelasticity in the P33 (top panel) and the P11 wave
(bottom panel).  Notation is as in Fig.~\ref{fig:delta3311}.} 
\label{fig:inel3311}
\end{figure}

In order to investigate the importance of different degrees 
of freedom we include in the first step only the $\pi N$ and
the $\pi\Delta$ channels without the background.
In this case the model reproduces the amplitudes at lower 
energies provided we take a value for $r_{N\Delta}$ which 
is substantially larger than unity (Fig.~\ref{fig:delta3311}).
When the background is included, the agreement considerably 
improves except for the energies close to the $\Delta(1600)$ 
resonance.
This is most notably seen in Fig.~\ref{fig:inel3311} in which
the inelasticity exhibits a qualitatively different behavior 
compared to the case with no background, and becomes consistent 
with the experimental data up to $W\approx 1700$~MeV.
A small kink around 1700~MeV is an indication of the
$\Delta(1600)$ resonance.

Including the $\sigma\Delta$ and the $\pi R$ channels 
gives an almost perfect fit to the experimental amplitudes also 
in the vicinity of the $\Delta(1600)$ resonance, 
washing out almost completely the signature of that 
resonance in the phase shift (Fig.~\ref{fig:delta3311}).

\begin{figure}[h!]
\includegraphics[width=8cm]{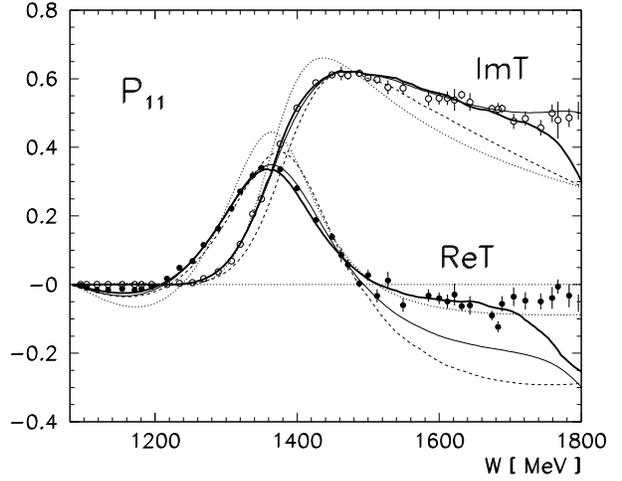}
\caption{The real and imaginary parts of the T matrix for
the P11 partial amplitudes.  Notation as in Fig.~\ref{fig:delta3311}.}
\label{fig:retimt11}
\end{figure}

Turning to the case of the P11 partial wave we note that including 
only the $\pi N$ and $\pi\Delta$ channels without the background
the Born approximation fails to reproduce the amplitudes determined 
in the partial wave 
analysis even if we considerably increase the values of some
coupling constants and take $r_{N R}=2.20$ and $r_{\Delta R}=1.55$ 
(Figs.~\ref{fig:delta3311} and \ref{fig:retimt11}).
Adding the background yields the correct behavior of the amplitudes
below the two-pion threshold. 
In order to reproduce the amplitudes in the region of the Roper 
resonance we have to keep the large value for $r_{\Delta R}$,
still, the approximation does not reproduce the rapid rise of 
the inelasticity just above the two-pion threshold nor 
the property that it remains close to unity even well above the 
Roper resonant energy (Fig.~\ref{fig:inel3311}).
Including the $\sigma N$ channel reproduces the threshold behavior 
and considerably improves the agreement at higher energies.

\begin{figure}[h!]
\includegraphics[width=8cm]{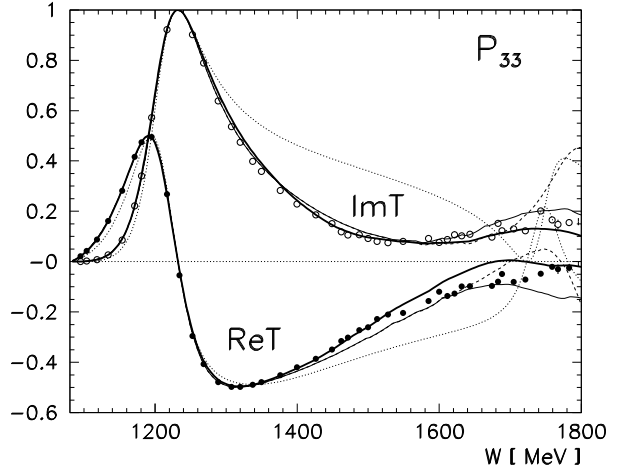}
\caption{The real and imaginary parts of the T matrix for
the P33 partial amplitudes.  Notation as in Fig.~\ref{fig:delta3311}.}
\label{fig:retimt33}
\end{figure}

Let us stress that in order to reproduce the widths of the 
$N(1440)$ and $\Delta(1232)$ resonances, the respective $\pi N R$ 
and $\pi N\Delta$ coupling constants have to assume considerably 
larger values than the corresponding bare-quark values.
The choice of the model parameters $r_{BB'}$ (\ref{BVB})
that drive these constants turns out 
to be quite different for the P33 and the P11 case. 
A possible solution to these inconsistencies is presented 
in the next section.

\subsection{\label{sec:results-full}Results of the full calculation}

The results presented here have been obtained using the same set of 
parameters as in the Born approximation for the bag radius, the mass 
and the width of the $\sigma$-meson, and the position of the K-matrix 
poles (see Table~\ref{tab:parameters}).
Similar results are obtained for $0.75~\mathrm{fm}<R<1.0~\mathrm{fm}$ 
as well as for $400~{\rm MeV} < m_\sigma < 550$~MeV provided the 
coupling constants are slightly readjusted.
We keep the same set of parameters for both partial waves;
we allow only for a slight deviation in the parameters entering 
the kernels.

\begin{figure}[h!]
\includegraphics[width=8cm]{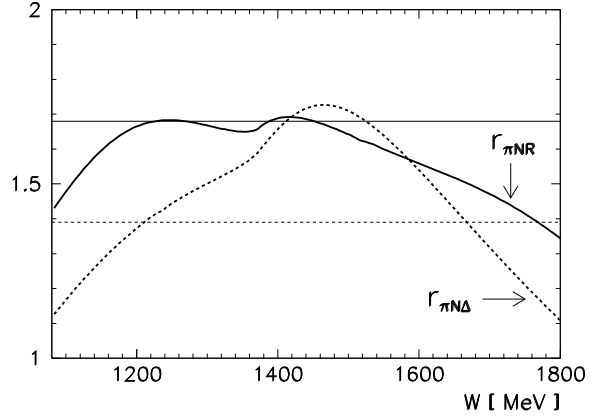}
\caption{Ratios (\ref{geff}) of the  $\pi N R$ (solid line)
and the $\pi N\Delta$ (dashed line) transition amplitudes  
to the respective bare quark values as a function of the 
invariant mass $W$. 
Straight lines: corresponding constant values used in the Born 
approximation.
Model parameters as in Table~\ref{tab:parameters}.}
\label{fig:gpiNX}
\end{figure}

We first investigate the relation between the bare
matrix elements of the baryon-meson interaction (\ref{BVB}) 
and the corresponding dressed values $\mathcal{V}_{BB'}$
which are solutions of the integral equations (\ref{eq4calVBN})
and (\ref{eq4calVBD}).
In the case of the elastic channel we introduce the ratio
\begin{equation}
r_{\pi NB}  = {\mathcal{V}_{NB}(k_0)\over V_{NB}(k_0)}
\label{geff}
\end{equation}
measuring the renormalization of the bare vertex.
In the case of the $N(1440)$ and the $\Delta(1232)$ we find that 
this ratio exhibits a relatively strong energy dependence 
(Fig.~\ref{fig:gpiNX}) and yields a substantial enhancement of 
the bare coupling constant over a broad energy range.
The enhancement is consistent with the value of the corresponding
coupling constant used in our analysis in the Born approximation.

The formulas (\ref{eq4calVBN}) and (\ref{eq4calVBD}) represent 
a system of coupled non-linear integral equations for the dressed 
vertices since the kernels (\ref{calKNN}) --  (\ref{calKDD})
themselves contain these vertices.
We have not attempted to solve the system exactly but have
substituted the dressed vertices appearing in the kernels by
$\mathcal{V}_{BB'}(k)=\bar{r}_{\pi BB'}V_{BB'}(k)$
with suitably chosen values for $\bar{r}_{\pi BB'}$.
We have adjusted these values by averaging the corresponding
solutions for the dressed vertices in the relevant energy range,
allowing for small variations to obtain better overall fits.
This approximation is justified because the contribution
from the integrals in (\ref{eq4calVBN}) and (\ref{eq4calVBD}) 
turns out to be less important compared to the leading term.  
Also, many of the dressed vertices only negligibly influence 
the result such that we can put the corresponding 
$\bar{r}_{\pi BB'}=1$.
In Table 1 only a few relevant cases are listed; 
in all other cases  $\bar{r}_{\pi BB'}=1$.

\begin{figure}[h!]
\includegraphics[width=8cm]{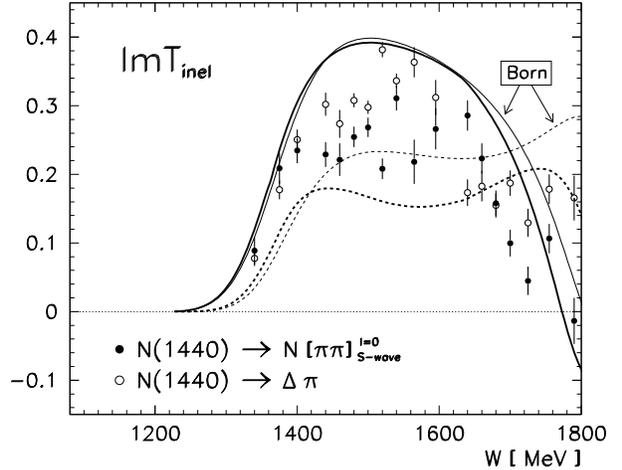}
\caption{Inelastic contributions to the imaginary part
of the T matrix for the P11 partial wave.  Full lines
and filled circles: contribution of
$N(1440)\to N[\pi\pi]^{I=0}_{s-\mathrm{wave}}$.
Dashed lines and empty circles: contribution of 
$N(1440)\to\Delta\pi$.
Thin lines represent the Born approximation.
The data are from \protect\cite{manley92}.
Model parameters as in Table~\ref{tab:parameters}.}
\label{fig:imtinel11}
\end{figure}

In the P11 wave the calculated amplitudes closely follow the 
experimental values in the energy range from the threshold 
up to $W\sim 1700$~MeV  
(Figs.~\ref{fig:retimt11} and \ref{fig:inel3311}).
In particular, we should stress the excellent agreement of
the amplitudes slightly above the $2\pi$ threshold where the
inelasticity is dominated by the $\sigma N$ channel.
Compared to the results of the Born approximation we notice
an improvement at very low energies as well as
at those above the resonance, which is a consequence
of the energy-dependent dressing.

Above $W\sim 1700$~MeV the imaginary part of the amplitude 
as well as the inelasticity drop rapidly. 
The simple model involving only the Roper-like bound state 
and the nucleon breaks down here and the effects of other 
resonances as well as other channels such 
as the $\rho N$ and $\eta N$ may become more important.
To investigate this point we have included in our calculation
the next P11 excitation, the $N(1710)$, and assumed that it 
couples only to the  $\sigma N$ channel, representative of a 
generic $\pi\pi$ decay with $\sim 80$~\% of the value 
of $G_\sigma$.
(The experimental branching fraction for 
$N(1710)\rightarrow N\pi\pi$ is (40--90)~\%.)
A better agreement at higher energies is obtained, 
although a similar effect can be achieved by decreasing
the bag radius (or increasing the cut-off parameter in general)
which makes a model-independent analysis less reliable.

In Fig.~\ref{fig:imtinel11} we display the imaginary parts
of the off-diagonal T-matrix elements corresponding to
the processes $\pi N\to\pi\Delta$ and $\pi N\to\sigma N$
calculated with the $N(1710)$ included.
The amplitudes are averaged over the range of unstable hadron
invariant masses as described in sect.~\ref{sec:average}.
In comparison to the values extracted from a recent partial-wave
analysis of $\pi N\to\pi N$ and $\pi N\to \pi\pi N$ \cite{manley92},
our model overestimates the $\sigma N$ decay probability
and underestimates the $\pi\Delta$ channel.
The disagreement is a consequence of the destructive interference
between the nucleon and the bare Roper in this channel at higher $W$. 
In the Born approximation, a better agreement in both inelastic
channels is obtained.  Had we departed from our standard set
of parameters chosen to yield a good overall description of
various amplitudes in both partial waves, or by
including further channels in the integral equations,
an improvement can be achieved in the full calculation as well.
(Note also the large systematic scatter of data.)

The full calculation for the P33 partial wave does not
significantly improve the results of the Born 
approximation which are anyway  satisfactory throughout 
the energy range from the threshold to $W\sim 1700$~MeV.
It does, however, explain the strong enhancement of the bare
quark-model $\pi N\Delta$ coupling of the Born approximation.
Above $W\sim 1800$~MeV the agreement is lost even if we
include another bare resonant state corresponding to
the $\Delta(1910)$, indicating the need to include further
channels not present in our analysis.

\begin{figure}[h!]
\includegraphics[width=8cm]{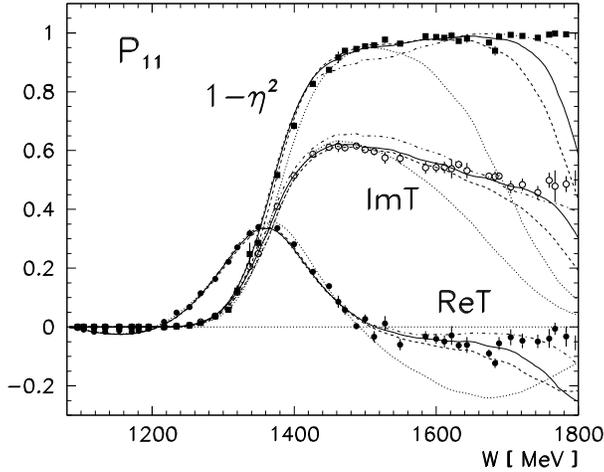}
\caption{The real and imaginary parts of the T matrix and 
the inelasticity for the P11 partial wave for different 
bag radii: $R=1$~fm (doted lines), $R=0.9$~fm (dashed lines), 
$R=0.83$~fm (solid lines), and $R=0.77$~fm (dashed-dotted lines).}
\label{fig:rdep11}
\end{figure}

In order to test the sensitivity of the results we have varied
the parameters appearing in Table~\ref{tab:parameters}.
Varying the bag radius and the parameters $r_{NB}$ we find 
that in the P33 case the results are almost independent of 
the bag radius while a few percent change in $r_{NB}$'s may
improve the agreement.
In the P11 case, the results are more sensitive to the choice
of the bag radius (Fig.~\ref{fig:rdep11}).
For $R\sim 0.8$~fm the agreement at larger energies 
improves; for $R\lesssim 0.75$~fm the real part of the amplitude
comes too close to zero and the phase shift loses its
characteristic resonant behavior.
The optimal value of $r_{N R}$ increases by $\sim 5$~\% for $R=1.0$~fm
and decreases by $\sim 10$~\% for $R=0.77$~fm.

In Table~\ref{tab:parameters} we list the positions of poles 
of the K matrix which are free parameters in our approach.
Alternatively, we could have used the masses of the bare
quark states entering our ansaetze for the PV states 
(\ref{PsiNa}),  (\ref{PsiBa}) and  (\ref{chi2K});
these masses are related to the pole values 
through the procedure described in sect. 3.5.
From the values of the bare masses we may obtain an indication
of how sensible is the comparison of baryon masses calculated 
in various quark models (with no mesons included) and the 
experimental positions of the corresponding resonances.
We are not able to give a conclusive answer 
since the result is strongly sensitive on some quantities 
(in particular the bag radius and the $\pi RR$ coupling constant)
which only weakly influence the behaviour of the scattering 
amplitudes and are therefore not well fixed in our calculation.
Nevertheless, we can conclude that the self-energy corrections
and the effects of resonance mixing most strongly affect 
the ground state, lowering its mass compared to the bare 
mass by 310~MeV at $R_\mathrm{bag}=0.83$~fm  (or by 180~MeV 
at $R_\mathrm{bag}=1.0$~fm). The bare delta-nucleon 
mass difference turns out to be smaller than the 
experimental one by about 90~MeV; the remaining difference 
can be attributed to the chromomagnetic interaction.
In the case of the Roper resonance the bare mass splitting 
is smaller by 120~MeV at $R_\mathrm{bag}=0.83$~fm (or by 60~MeV at 
$R_\mathrm{bag}=1.0$~fm).  Since our value for the K matrix 
pole is typically 100 MeV higher than the experimental value,
the resulting bare Roper-nucleon mass splitting remains in 
the ballpark of admissible values. A similar conclusion holds 
for the $\Delta(1600)$ where the bare mass splitting is 
typically 200~MeV lower than the value deduced from the 
K matrix pole.

\section{\label{sec:conclusions}Conclusions and perspectives}

  We have developed a general method to incorporate the
excited baryons represented as quasi-bound  quark-model states
into a dynamical framework with correct unitarity and symmetry
requirements and proper boundary conditions.
To illustrate the key points, we have used the Cloudy Bag Model, 
although the method is applicable to more sophisticated models.
In fact, the only information needed as input to our computational
scheme are the matrix elements of the meson interaction between
the quark states, {\it i.e.\/} $\langle\Phi(k)|V|\Phi(k=0)\rangle$.
In a more ambitious computational scheme, the method can
be extended to allow for the readjustment of the intrinsic
three-quark wave-function to the scattering boundary 
conditions by using the Kohn variational principle.

We have shown that an intricate interplay of the $\pi N$, 
$\pi\Delta$ and $\sigma N$ degrees of freedom governs the 
elastic and inelastic pion-nucleon scattering in the energy 
range from the threshold up to $W\sim 1700$~MeV.
The model 
explains at least qualitatively the behavior of the amplitudes 
in the vicinity of the $\Delta(1600)$  which can be regarded 
as the P33-wave counterpart of the $N(1440)$ in the P11 wave.
We have studied in a systematic way the role of the background 
processes which turn out to be important to qualitatively 
reproduce the experimental data throughout the first and 
the second resonance region.

We have described the correlated two-pion decay in the 
relative $s$-wave by the $\sigma$-meson.
In spite of this purely phenomenological approach, our results 
show that this degree of freedom is crucial to explain certain 
features of the scattering amplitudes, in particular the 
inelasticity for the P11 partial wave just above the two-pion 
threshold which rapidly rises from zero to unity and remains 
close to this value in a broad energy range.
Because of the $s$-wave nature of the $\sigma$-meson coupling 
compared to the $p$-wave coupling of the pion to the $\Delta$,
the two-pion decay is dominated by the $\sigma N$ channel
in the energy region slightly above the two-pion threshold.
This offers the possibility to determine the strength of
the coupling in a model-independent way, which is generally 
not the case at higher energies where the same level of 
agreement can be obtained in a broad range of parameters.
Our calculation indicates that a low mass of the $\sigma$-meson
($m_\sigma\approx450$~MeV) is preferable over the larger mass 
typically obtained in partial-wave analyses.

We have designed a framework to numerically solve the integral 
equation for the K matrix by approximating the kernels with 
a separable form which preserves the symmetries of the matrix
and thus ensures unitarity.
The important outcome of this calculation is a substantial
increase of the quark-model pion-baryon couplings explaining the 
large width of the Roper and possibly other resonances.
Such a strong enhancement could --- at least partially ---
accommodate relatively weak pion-baryon coupling strengths
predicted in constituent quark models.
This shows, on the one hand, that one can explain the nature
of the Roper resonances without invoking ``exotic'' degrees
of freedom mentioned in the Introduction and, on the other hand,
establishes one of the benchmarks for an assessment
of the underlying quark models. 
Still, the scattering analysis alone should not be expected
to provide a definitive selection criterion, because 
a consistent description, as shown in the example above,
can be achieved within a relatively broad range of parameters.
A future analysis devoted to pion electro-production
could provide a more complete set of criteria. 
The application of the method to calculate the electro-production
amplitudes will be treated in a separate paper.


\appendix
\section{\label{aaPsi}Evaluation of one- and two-pion matrix elements}

We derive some expressions for the matrix elements of the 
pion field between the eigenstates of the Hamiltonian with the
pion part given by (\ref{Hpi}). 
If $\Psi_A$ is an eigenstate then
\begin{equation}
    (\omega_k+H-E_A)a_{mt}(k)|\Psi_A\rangle   =  
  - V^\dagger_{mt}(k)|\Psi_A\rangle\,,
\label{commute1}
\end{equation}
\begin{eqnarray}
 &&  \hspace{-15pt}(\omega_k+\omega_k'+H-E_A) a_{mt}(k)a_{m't'}(k')
|\Psi_A\rangle 
\nonumber \\ &&
=  
  - \left(V^\dagger_{mt}(k) a_{m't'}(k') +
    V^\dagger_{m't'}(k')a_{mt}(k)\right)|\Psi_A\rangle\,,\qquad
\label{commute2a}
\end{eqnarray}

Multiplying (\ref{commute1}) by an eigenstate of (\ref{Hpi})
we obtain
\begin{eqnarray}
\langle\widehat{\Psi}_B(k)|  a_{mt}(k)|\Psi_A\rangle   
&=&  
\nu\delta(\omega_k+E_B(k)-E_A)
\nonumber\\ &&
  - {\langle\widehat{\Psi}_B(k)|V^\dagger_{mt}(k)|\Psi_A\rangle
    \over  (\omega_k+E_B(k)-E_A)}\,,
\label{commute1a}
\end{eqnarray}
where $\nu$ is an arbitrary constant; it is used to determine 
the normalization of the channel states (\ref{PsiN}),
(\ref{PsiBa}) and (\ref{Psisia}).

Taking for $\Psi_A$ either (\ref{PsiN}) or (\ref{PsiBa}), 
multiplying (\ref{commute2a}) by an eigenstate with momentum 
$\vec{k}+\vec{k}'$, and neglecting the terms with two or 
more pions, we obtain
\begin{eqnarray}
 &&  \hspace{-15pt}
(\omega_k+\omega_k'+E_{B}(\vec{k}+\vec{k}')-W) 
\langle\widehat{\Psi}_B(\vec{k}+\vec{k}')|a_{mt}(k)
|\Psi_{B'}(k')\rangle 
\nonumber \\ &&
=  
  - \langle\widehat{\Psi}_B(\vec{k}+\vec{k}')|V^\dagger_{mt}(k) 
         |\Psi_{B'}(k')\rangle\,. 
\label{commute2b}
\end{eqnarray}
Note that when $B'$ is on-shell, $E_{B'}=W-\omega_k'$,
and (\ref{commute2b}) reduces to (\ref{commute1a}).


\section{\label{sec:wD}Derivation of the mass distribution function}

We derive here the expression for the mass distribution function 
$w_B(M)$ in (\ref{tildePsiD}) for the case of the $\pi\Delta$
channel which dominates the two-pion decay through the intermediate
baryon.
In this case we can assume that the intermediate $\Delta$
decays only into a pion and the nucleon.
(This assumption is justified since experimentally the elastic 
channel remains the dominant process also well above the two-pion 
threshold.)
Then the K matrix becomes a scalar (denoted as $K_\Delta$) and the 
orthonormalized state in (\ref{orthonormalization}) assumes the form
\begin{eqnarray}
 |\widetilde{\Psi}_\Delta(M)\rangle
&=& {1\over\sqrt{1 +K_\Delta(M)^2}}
    \sqrt{\omega_2M_N\over k_2M}\,c_\Delta^N(M)\,{1\over z_\Delta}
|\widehat{\Psi}_\Delta(M)\rangle
\nonumber \\ 
&\equiv&
   w_\Delta(M)|\widehat{\Psi}_\Delta(M)\rangle\,.
\label{tildePsiDa}
\end{eqnarray}
where $M$ is the invariant mass of the intermediate $\Delta$, while  
$\omega_2$ and $k_2$ are the energy and momentum of the second pion.
From (\ref{cal2chiN}) we see that close to the resonance
the pion amplitude behaves as
$\chi^{NN}_\Delta(k,k_2) \approx -c^N_\Delta(M)\,\mathcal{V}_{N\Delta}(k)$ 
and from (\ref{cal2chiBB}) we find 
$\widehat{\chi}^{BN}_\Delta(k,k_2,M_B) \approx - c^N_\Delta(M)\,
\mathcal{V}_{B\Delta}^{M_B}(k)$,
with $c_\Delta^N\propto(M-M_\Delta)^{-1}$. 
The state  $\widehat{\Psi}_\Delta(M)$ introduced in (\ref{tildePsiDa})
can then be written in the form
\begin{eqnarray}
  |\widehat{\Psi}_\Delta(M)\rangle &=& 
z_\Delta\Biggl\{|\Phi_\Delta\rangle
\nonumber \\ && 
   - \int\d k\,{{\cal{V}}_{N\Delta}(k)\over\omega_k+E_N(k)-M}\,
      [a^\dagger(k)|\Psi_N\rangle]^{\th\th}
\nonumber\\ && 
   - \sum_B\int\d k\,{{\cal{V}}_{B\Delta}^{M_B}(k)
           \over\omega_k+E_B(k)-M}\,
      [a^\dagger(k)|\widehat{\Psi}_B\rangle]^{\th\th}\Biggr\}\,.
\nonumber\\ 
\label{hatPsiD}
\end{eqnarray}
Note that in the weak-coupling limit the form (\ref{hatPsiD})
corresponds to the usual perturbative expression for the 
$\Delta$ state, with $\Psi_N$ and  $\widehat{\Psi}_\Delta$ 
replaced by the corresponding bare quark states.
It is dominated by the bare-quark state $\Phi_\Delta$.
This is another reason for choosing the particular form of
factorization in (\ref{tildePsiDa}).

From (\ref{sol4chi}) and (\ref{chi2K}) we have 
($K_\Delta\equiv K_{NN}^{\th\th}$)
\begin{equation}
   K_\Delta(M) 
    =\pi\,{\omega_2M_N\over k_2M}\,
           {\mathcal{V}_{N\Delta}(k_2)^2
                            \over Z_\Delta(M)(M_\Delta - M)}
    + \ldots 
\label{calV2K}
\end{equation}
for $M\approx M_\Delta$, where the terms denoted by $\ldots$ vanish 
at the resonance; $Z_\Delta(W)$ is defined in (\ref{defZ}).
From (\ref{tildePsiDa}) it then follows
$$
    w_\Delta(M)
   = {K_\Delta\over z_\Delta\,\sqrt{1 +K_\Delta^2}}\,
     {1\over\pi} \sqrt{k_2M\over\omega_2M_N}
     {1\over\mathcal{V}_{N\Delta}(k_2)}\,. 
$$

At this point we can use the expression for the K matrix obtained
numerically or use a suitable parameterization either of the computed
form or of the experimental data. 
The simplest choice is 
\begin{equation}
    K_a = {C\over M_\Delta - M}\,,
\qquad
    C  = \pi\,{\omega^\Delta_2M_N\over k^\Delta_2M_\Delta}\,
           {\mathcal{V}_{N\Delta}(k^\Delta_2)^2
                                    \over Z_\Delta(M_\Delta)}\,,
\label{KDapprox}
\end{equation}
where $\omega_2^\Delta=(M_\Delta^2-M_N^2+m_\pi^2)/2M_\Delta$ and
the residue $C$ (corresponding to $\half\Gamma$) is assumed to be 
$W$-independent; the second equality follows from (\ref{calV2K}).
From  (\ref{tildePsiD}) and (\ref{hatPsiD}) we have
\begin{eqnarray}
    w_\Delta(M)
  &=& {1\over\sqrt{\pi C}}\,{K_a\over\sqrt{1 +K_a^2}}\,
     \sqrt{k_2\omega^\Delta_2M\over\omega_2k^\Delta_2M_\Delta}\,
      {\mathcal{V}_{N\Delta}(k^\Delta_2)
               \over \mathcal{V}_{N\Delta}(k_2)}\,
\nonumber \\ && \times
      {1\over z_\Delta\,\sqrt{Z_\Delta(M_\Delta)}}  \,.
\label{sol4wD}
\end{eqnarray}
For sufficiently small values of $C$, $w_\Delta(M)$ in (\ref{sol4wD}) is 
strongly peaked around $M_\Delta$ and has a unit integral
provided that $z_\Delta\,\sqrt{Z_\Delta(M_\Delta)}=1$ , hence
\begin{equation}
w_\Delta(M)^2 \rightarrow \delta(M-M_\Delta)\,,
\qquad
   C \rightarrow 0\,. 
\label{w2delta}
\end{equation}
In this limit we obtain the usual wave-function normalization of 
(\ref{hatPsiD}), {\it i.e.\/} $z_\Delta = Z_\Delta(M_\Delta)^{-1/2}$.

In a more precise calculation we use a better approximation for 
the K matrix, $K_a = C/(M_\Delta - M) + D$,
which turns out to give a very good approximation
for $M$ from the threshold to values well above the resonance.
Note, however, that the parameter $D$ {\em should not be identified\/}
with the background contribution since in the approximate formula 
(\ref{KDapprox}) the coefficient $C$ is kept fixed while in 
(\ref{calV2K}) all terms exhibit strong $k_2$ (or, equivalently, $M$) 
dependence.
The coefficient $D$ is chosen such as to compensate this dependence.
In our calculation we use $C=55$~MeV and for $D$ a constant value 
$-0.41$  below $M\sim 1400$~MeV and, above it, a value that smoothly 
approaches zero.


\section{\label{kernels}Evaluating the kernels}

To evaluate the kernel (\ref{calKNBdef}) in the integral equation 
for $\chi_{JT}$ amplitudes 
we insert  a complete set of states
$$
1= |\Psi_N\rangle\langle\Psi_N| 
  + \sum_B\,|\widetilde{\Psi}_B(M)\rangle
                \langle\widetilde{\Psi}_B(M)|\,.
$$
where the sum implies also the integral over invariant masses
and momenta.
Next we use the (adjoint of) (\ref{commute1a}) 
and (\ref{commute2b}).
Following Ericson and Weise (see \cite{Weise}, sect. 2.5.3.)
we substitute the momenta of the intermediate baryon by a suitably 
chosen average over $\vec{k} +\vec{k}'$ denoted as $\bar{k}$.  
The kernel then takes the form:
\begin{eqnarray*}
\mathcal{K}^{NB}_M(k,k') &=& f_{NB}^N\,
        {\langle\widehat{\Psi}_B||V(k)||\Psi_N\rangle
        \langle\Psi_N||V(k')||\Psi_N\rangle
                    \over\omega_k+\omega_k'+E_N(\bar{k})-W}
\nonumber\\ && 
+ \sum_{B'} f_{NB}^{B'}\int \d M'\,
  {\langle\widehat{\Psi}_{B}||V(k)||\widetilde{\Psi}_{B'}(M')\rangle
           \over\omega_k+\omega_k'+ E'(\bar{k}) - W}
\\ && \times
   \langle\Psi_N||V(k')||\widetilde{\Psi}_{B'}(M')\rangle\,,
\end{eqnarray*}
where 
\begin{eqnarray}
  f_{AB}^C  &=& \sqrt{(2J_A+1)(2J_B+1)(2T_A+1)(2T_B+1)}
\nonumber \\
&& \times
   W(1J_AJ_B1;J_C,J)W(1T_AT_B1;T_C,T)\,.
\label{fabc}
\end{eqnarray}
The matrix elements are those entering the expression
for the $\chi$'s in (\ref{cal2chiN})-(\ref{cal2chiBB}),
$
   \langle\Psi_N||V(k')||\widetilde{\Psi}_{B'}(M')\rangle
  =
 w_{B'}(M')\break \mathcal{V}_{B' N}(k')
$
and 
$
\langle\widehat{\Psi}_B(M)||V||\widetilde{\Psi}_{B'}(M')\rangle
=
 w_{B'}(M')\mathcal{V}_{B'B}^M
$.
We can now approximately perform the integration over $M'$
assuming (\ref{w2delta}):
\begin{eqnarray*}
\mathcal{K}^{NB}_M(k,k') &\approx&  
      f_{NB}^N\,{\mathcal{V}_{NB}^M(k)\mathcal{V}_{NN}(k')
       \over\omega_k+\omega_k'+E_N(\bar{k})-W}
\\ &&
   + \sum_{B'} f_{NB}^{B'}
         \,{\mathcal{V}_{B' B}^M(k)\mathcal{V}_{B' N}(k')
            \over \omega_k+\omega_k'+ E_{B'}(\bar{k}) - W} + \ldots
\end{eqnarray*}
In order to build up a feasible computational scheme we 
approximate the kernel with a separable expression, a relativistic 
extension of the approximation used in Ref.~\cite{thomas8180}:
\begin{eqnarray*}
&&   {1 \over \omega_k+\omega_k'+E_{B'}(\bar{k})-W}
\nonumber \\ &&
\approx 
    {\omega_0 + \omega_1 + E_{B'}(\bar{k}) - W
     \over(\omega_k'+E_{B'}(\bar{k})-E_N(k_0))
          (\omega_k+E_{B'}(\bar{k})-E(k_1))}\,.
\nonumber 
\end{eqnarray*}
Here $k_1$, $\omega_1$ are the on-shell pion momentum and 
energy in the $\pi B$ channel satisfying
$W = \omega_1 + E(k_1) = \omega_0 + E_N(k_0)$.
The approximation on the RHS 
coincides with the exact expression on the LHS  when either
of the two pions is on-shell, i.e when either
$\omega_k=\omega_0$ or $\omega_k'=\omega_1$.
(This can be easily seen by writing  $W=\omega_0 + E_N(k_0)$
on the LHS and $W = \omega_1 + E(k_1)$ on the RHS in
the first case and vice versa in the second one.)
When both pions are on-shell, the denominator can be cast in the form
\begin{equation}
 {1\over \omega_1 +  E_{B'}(\bar{k}) -E_N(k_0)}
\approx
{E_{B'}(\bar{k}) + E_N(k_0)-\omega_1 \over
    M_{B'}^2 + 2\omega_1E_N(k_0) - M_N^2 - m_\pi^2}\,. 
\label{udenom}
\end{equation}
We have assumed $\langle\vec{k}_0\cdot\vec{k}_1\rangle=0$ which 
is essentially the same approximation suggested in \cite{Weise}
and the denominator acquires the characteristic $u$-channel form.
Since we describe the resonance in terms of the wave function rather 
than bispinors, our numerator differs from the correct relativistic 
expression in the $u$-channel (see {\it e.g.\/} \cite{Weise}).
Since our expression is anyway approximate, we replace the factor 
$E_{B'}(\bar{k})+E_N(k_0)-\omega_1=E_{B'}(\bar{k})+E_{B}(k_1)-\omega_0$ 
in our numerator by the correct relativistic expression $2M_{B'}$.

In the general case we find:
$$
  {\cal K}_{MM'}^{BB'}(k,k')
=  
  \sum_{B''} f_{BB'}^{B''}
  {\mathcal{V}_{B''B}^{M}(k') \, \mathcal{V}_{B''B'}^{M'}(k)
   \over \omega_k+\omega_k'+E_{B''}(\bar{k})-W}\,,
$$
where the sum includes also the ground states.
We now approximate
\begin{eqnarray*}
&&\hspace{-24pt}
   {1 \over \omega_k+\omega_k'+E_{B''}(\bar{k})-W}
\nonumber \\
&&
\hspace{-24pt}
\approx {\omega_1 + \omega_1' + E_{B''}(\bar{k}) - W
     \over(\omega_k'+E_{B''}(\bar{k})-E (k_1 ))
          (\omega_k +E_{B''}(\bar{k})-E'(k_1'))}\,.
\label{approx2poleD}
\end{eqnarray*}
Here $k_1$, $\omega_1$,  $k_1'$, $\omega_1'$ are the on-shell
pion momenta and energies satisfying
$
    W = \omega_1 + E(k_1) = \omega_1' + E'(k_1')
$,
$
   E(k_1) = \sqrt{M^2+k_1^2}
$,
$
   E'(k_1') = \sqrt{M'{}^2+k_1'{}^2}
$.
The approximation on the RHS of (\ref{approx2poleD}) coincides
with the exact expression on the LHS of (\ref{approx2poleD}) 
when either pion is on the mass shell, {\it i.e.\/} when either
$\omega_k=\omega_1'$ or $\omega_k'=\omega_1$.
When on-shell, the kernels are proportional to the $u$-channel 
background elements of the K matrix.
We now use the same approximation as in (\ref{udenom}) for $\bar{k}$
as well as for the numerator;
the final forms are given by (\ref{calKNN})--(\ref{calKDD}).

In the case of the channels involving the $\sigma$-meson we find
\begin{eqnarray*}
{\cal K}^{B\sigma}_{M\mu}(k,k') &=&
    \sum_{B'}
   {V^\mu_{B'B}(k')V_{B'N}(k)\over 
     \omega_{\mu k}' + \omega_k + E_{B'}(\bar{k}) - W}
\nonumber\\
    &=& {\cal K}^{\sigma B}_{\mu M}(k',k)\,, 
\nonumber\\
{\cal K}^{\sigma\sigma}_{\mu\mu'}(k,k') &=&  
    \sum_{B'}
    {V^{\mu'}_{B'N}(k')V^\mu_{B'N}(k)\over 
     \omega_{\mu k}' + \omega_{\mu k} + E_{B'}(\bar{k}) - W}\,.
\kern12pt
\end{eqnarray*}
where in the P11 case the sum over $B'$ includes
only $J=T=\half$ isobars; in the P11 case
the $N$ in the above expression is replaced by $\Delta$ while
the sum over $B'$ includes only $J=T=\thalf$ isobars.

\end{document}